\documentclass[%
 reprint,
 superscriptaddress,
 bibnotes,
 amsmath,amssymb,
 aps,
 prb,
 floatfix,
]{revtex4-2}

\usepackage{xspace}
\usepackage{graphicx}
\usepackage{dcolumn}
\usepackage{bm}
\usepackage{hyperref}
\usepackage{float}
\usepackage{ulem}
\usepackage[T1]{fontenc}
\usepackage{color}
\hypersetup{
  colorlinks   = true,
  urlcolor     = black,
  linkcolor    = black,
  citecolor   = blue
}
\usepackage{etoolbox}
\apptocmd{\sloppy}{\hbadness 10000\relax}{}{}
\usepackage{comment}

\usepackage{xcolor}

\newcommand{\bL}{\mathbf{L}}

\def\bL{\mathbf{L}}

\begin{document}

\title{Nanoscale imaging of spin textures \\with locally varying altermagnetic response in $\alpha$-Fe$_2$O$_3$}

\author{R. Yamamoto}
\thanks{These authors contributed equally to this work.}
\affiliation{Max Planck Institute for Chemical Physics of Solids, 01187 Dresden, Germany}
\affiliation{International Institute for Sustainability with Knotted Chiral Meta Matter (WPI-SKCM$^{2}$), Hiroshima University, Hiroshima 739-8526, Japan}

\author{S. Mayr}
\thanks{These authors contributed equally to this work.}
\affiliation{Paul Scherrer Institut, 5232 Villigen PSI, Switzerland}
\affiliation{Laboratory for Mesoscopic Systems, Department of Materials, ETH Z\"urich
, 8093 Z\"urich, Switzerland}

\author{A. Hariki}
\affiliation{Department of Physics and Electronics, Graduate School of Engineering,
Osaka Metropolitan University, 1-1 Gakuen-cho, Nakaku, Sakai, Osaka 599-8531, Japan}

\author{S. Finizio}
\affiliation{Paul Scherrer Institut, 5232 Villigen PSI, Switzerland}

\author{K. Sakurai}
\affiliation{Department of Physics and Electronics, Graduate School of Engineering,
Osaka Metropolitan University, 1-1 Gakuen-cho, Nakaku, Sakai, Osaka 599-8531, Japan}

\author{E. Weschke}
\affiliation{Helmholtz-Zentrum Berlin f\"ur Materialien und Energie GmbH, 12409 Berlin, Germany}

\author{K. Litzius}
\affiliation{Universität of Augsburg, 86150 Augsburg, Germany}

\author{M. T. Birch}
\affiliation{RIKEN Center for Emergent Matter Science (CEMS), Wako, Saitama 351-0198, Japan}

\author{L. A. Turnbull}
\affiliation{Max Planck Institute for Chemical Physics of Solids, 01187 Dresden, Germany}
\affiliation{Diamond Light Source, Didcot, Oxfordshire, OX11 0DE, United Kingdom}

\author{E. Zhakina}
\affiliation{Max Planck Institute for Chemical Physics of Solids, 01187 Dresden, Germany}

\author{M. Di Pietro Mart\'inez}
\affiliation{Max Planck Institute for Chemical Physics of Solids, 01187 Dresden, Germany}
\affiliation{International Institute for Sustainability with Knotted Chiral Meta Matter (WPI-SKCM$^{2}$), Hiroshima University, Hiroshima 739-8526, Japan}

\author{J. Reuteler}
\affiliation{ETH Z\"urich, 8093 Z\"urich, Switzerland}

\author{F. Schulz}
\affiliation{Max-PLanck-Institut f\"ur Intelligente Systeme, 70569 Sttutgart, Germany}

\author{M. Weigand}
\affiliation{Helmholtz-Zentrum Berlin f\"ur Materialien und Energie GmbH, 12409 Berlin, Germany}

\author{J. Raabe}
\affiliation{Paul Scherrer Institut, 5232 Villigen PSI, Switzerland}

\author{G. Sch\"utz}
\affiliation{Max-PLanck-Institut f\"ur Intelligente Systeme, 70569 Sttutgart, Germany}

\author{S. S. P. K. Arekapudi}
\affiliation{Helmholtz-Zentrum Dresden-Rossendorf, 01328 Dresden, Germany}
\affiliation{Technische Universität Chemnitz, 09111 Chemnitz, Germany}

\author{O. Hellwig}
\affiliation{Helmholtz-Zentrum Dresden-Rossendorf, 01328 Dresden, Germany}
\affiliation{Technische Universität Chemnitz, 09111 Chemnitz, Germany}

\author{W. H. Campos}
\affiliation{Max Planck Institute for the Physics of Complex Systems, 01187 Dresden, Germany}

\author{L. \v{S}mejkal}
\affiliation{Max Planck Institute for the Physics of Complex Systems, 01187 Dresden, Germany}

\author{J. Kune\v{s}}
\affiliation{Department of Condensed Matter Physics, Faculty of Science, Masaryk University, Kotl\'{a}\v{r}sk\'{a}  2, 61137 Brno, Czech Republic}

\author{C. Donnelly}
\affiliation{Max Planck Institute for Chemical Physics of Solids, 01187 Dresden, Germany}
\affiliation{International Institute for Sustainability with Knotted Chiral Meta Matter (WPI-SKCM$^{2}$), Hiroshima University, Hiroshima 739-8526, Japan}

\author{S.~Wintz}
\affiliation{Helmholtz-Zentrum Berlin f\"ur Materialien und Energie GmbH, 12409 Berlin, Germany}

\date{\today}

\begin{abstract}
Altermagnetism is a recently identified magnetic state in which time-reversal symmetry is broken despite a collinear compensated spin structure. The response of altermagnets is determined not only by their $d$-, $g$-, or $i$-wave spin order, but also the orientation of their N\'eel vector $\mathbf{L}$. Therefore, accessing a response that fundamentally depends on the orientation of $\mathbf{L}$ remains experimentally challenging in particular at the nanoscale. Here, we harness nano-spectroscopic X-ray magnetic circular dichroism (XMCD) to investigate nanoscale modulated altermagnetic responses in $\alpha$-Fe$_2$O$_3$ (Hematite). By performing spectroscopy across the temperature-induced $\mathbf{L}$-reorientation Morin transition, we observe the on-and-off switching of XMCD, in agreement with our theoretical calculations. Although the bulk XMCD vanishes below the Morin temperature, we confirm the reorientation of $\mathbf{L}$ by harnessing polarization-independent X-ray absorption spectroscopy.
Moreover, we observe a finite XMCD signal in nanoscale domain walls below the Morin transition. The domain walls host locally modulated N\'eel vectors and therefore exhibit a local altermagnetic response, while the surrounding domains show no XMCD.
At room temperature, we instead identify altermagnetic meron spin textures that exhibit XMCD in their planar regions but no XMCD in their nanoscopic cores. Our results establish a pathway to harness complex spin textures with nanoscale functionalities in a broader class of altermagnets with various $\mathbf{L}$-orientations and using light, earth-abundant elements.
\end{abstract}

\maketitle


\begin{figure*}[h!t]
\begin{center}
\includegraphics[width=1.8\columnwidth]{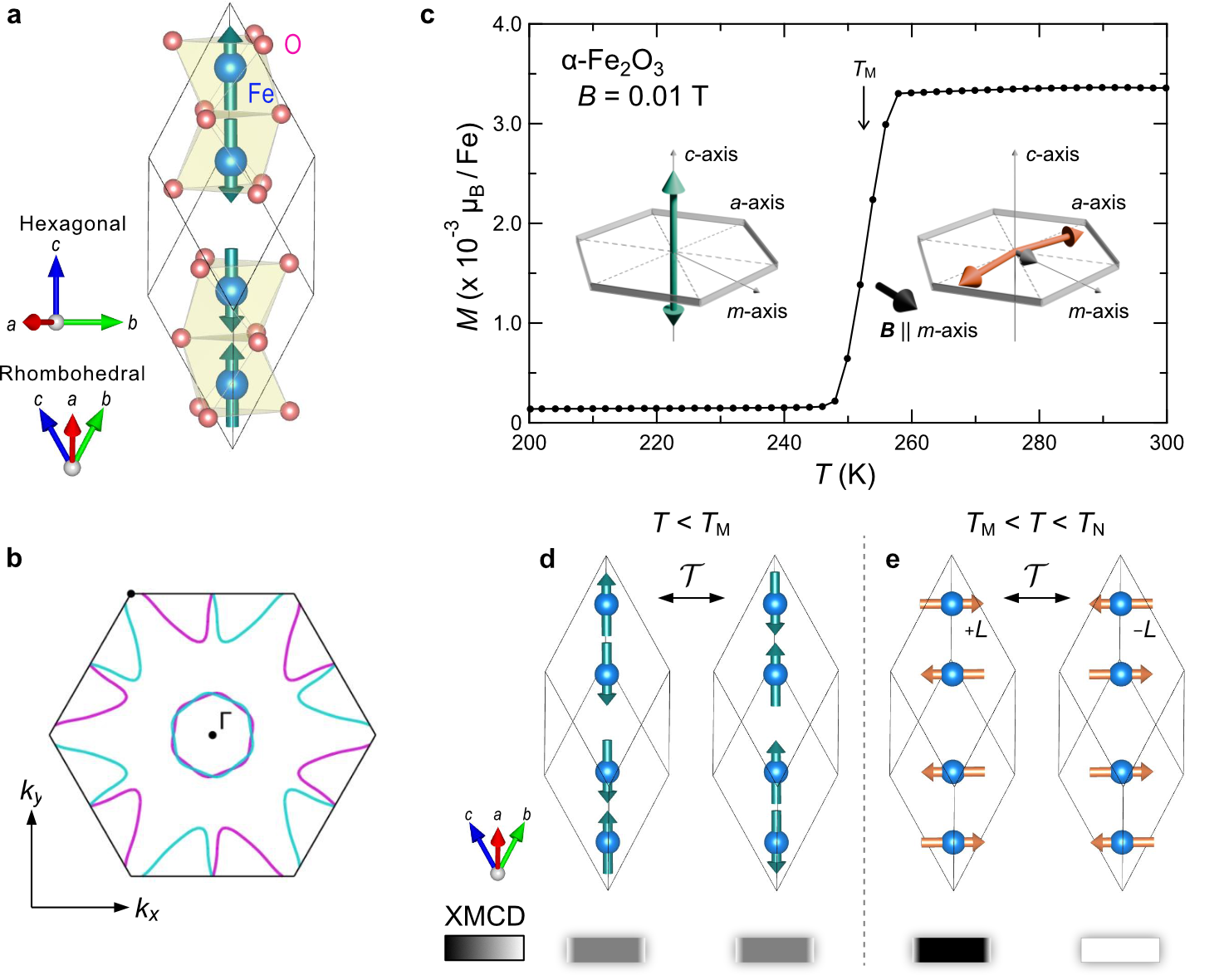}
\end{center}
\caption{\textbf{a} Magnetic structure of $\alpha$-Fe$_2$O$_3$ in a rhombohedral unit cell. The hexagonal crystallographic axes are also indicated in the figure.
\textbf{b} Altermagnetic spin-polarization calculated in momentum space for the $k_z=0$ plane and an energy of 198.5 meV below the valence band maximum.
\textbf{c} Temperature dependence of the magnetization of Sample \#1 measured under a magnetic field of 0.01 T along the hexagonal $m$-axis. N\'eel vectors in a hexagonal setting below and above $T_\mathrm{M}$ are schematically illustrated in the inset. Magnetic moments (orange arrows) above $T_\mathrm{M}$ are slightly canted within the $c$-plane, resulting in a small uncompensated moment (gray arrow). The uncompensated magnetic moment is oriented by the magnetic field, resulting in the N\'eel vector becoming perpendicular to the magnetic field. Below $T_\mathrm{M}$, the magnetic moments (green arrows) align parallel to the $c$-axis without canting.
\textbf{d,e} Magnetic structure of $\alpha$-Fe$_2$O$_3$ in a rhombohedral unit cell below and  above $T_\mathrm{M}$. Under time-reversal symmetry operation $\mathcal{T}$, the N\'eel vector $\bf{L}$ reverses its orientation. Expected X-ray magnetic circular dichroic (XMCD) contrast for normal-incidence X-rays $\mathbf{k} \parallel c$-axis for each spin configuration is schematically illustrated in the bottom.
}
\label{fig_1}
\end{figure*}

Altermagnets are collinear magnets which exhibit additional time reversal symmetry $\mathcal{T}$ breaking associated with the local symmetry of each sublattice~\cite{Smejkal22a,Smejkal22b}.
The recent proposal of altermagnetism has triggered extensive research not only into its fundamental physical properties, but also into potential device applications.
Following the theoretical characterization of altermagnets by non-relativistic, exchange-driven $d$-, $g$-, or $i$-wave type spin ordering, featuring two, four, or six spin-degenerated nodal surfaces in their non-relativistic band structure~\cite{Smejkal22a,Smejkal22b}, altermagnetic spin splitting was experimentally observed by angle-resolved photoemission spectroscopy~\cite{Krempasky24,Lee24, Osumi24,Reimers24,Zeng24,Ding24,Yang25,Li25,Lu25,Santhosh25}.

Beyond the spin splitting in momentum space, emergent phenomena arising from broken time-reversal symmetry despite antiferromagnetic ordering--such as the anomalous Hall effect (AHE)~\cite{Smejkal20,Feng22,Tschirner23,Wang23,Jeong25,Chu25,GonzalezBetancourt23,Kluczyk24,GonzalezBetancourt24,Reichlova24,Han24}, chiral magnons~\cite{Smejkal23,Liu24,Biniskos25}, and optical dichroic effects, including the magneto-optical Kerr effect (MOKE)~\cite{Marzin21,Iguchi25,Luo26,Pan26} and X-ray magnetic circular dichroism (XMCD)~\cite{Hariki24,Amin24,Yamamoto25,Takegami25,Galindez25}--have been reported. 
These phenomena require relativistic spin--orbit coupling (SOC), rendering them dependent not only on the type of altermagnet, but also on the orientation of the N\'eel vector $\mathbf{L} = \mathbf{m}_{1} - \mathbf{m}_{2}$, where $\mathbf{m}_{1,2}$ denote the magnetic moments of the two sublattices. 
This $\mathbf{L}$ orientation dependence is exemplified by the $g$-wave altermagnets $\alpha$-MnTe and CrSb. Although both are classified as $g$-wave altermagnets, altermagnetic responses such as AHE and XMCD are allowed in $\alpha$-MnTe with in-plane basal $\mathbf{L}$-orientation, whereas they are forbidden in CrSb exhibiting a perpendicular to the basal plane $\mathbf{L}$-orientation. This dependence of the physical properties of altermagnets  on the N\'eel vector opens the possibility for a spin reorientation-induced \textit{on/off switching} of emergent effects.
Thus, whether the response vanishes or remains finite goes beyond identifying the altermagnetic nature of materials and establishes a platform for controlling altermagnetic functionalities. 
Experimentally, recent studies demonstrating that a modulation of the emergent AHE can be realized through the reorientation of the N\'eel vector by application of strain and high magnetic fields~\cite{Feng22,Zhou25,Takagi25}.
However, while macroscopic spin reorientation enables global control of altermagnetic responses, achieving locally varying, nanoscale responses remains a major challenge.

\begin{figure*}[h!t]
\begin{center}
\includegraphics[width=2\columnwidth]{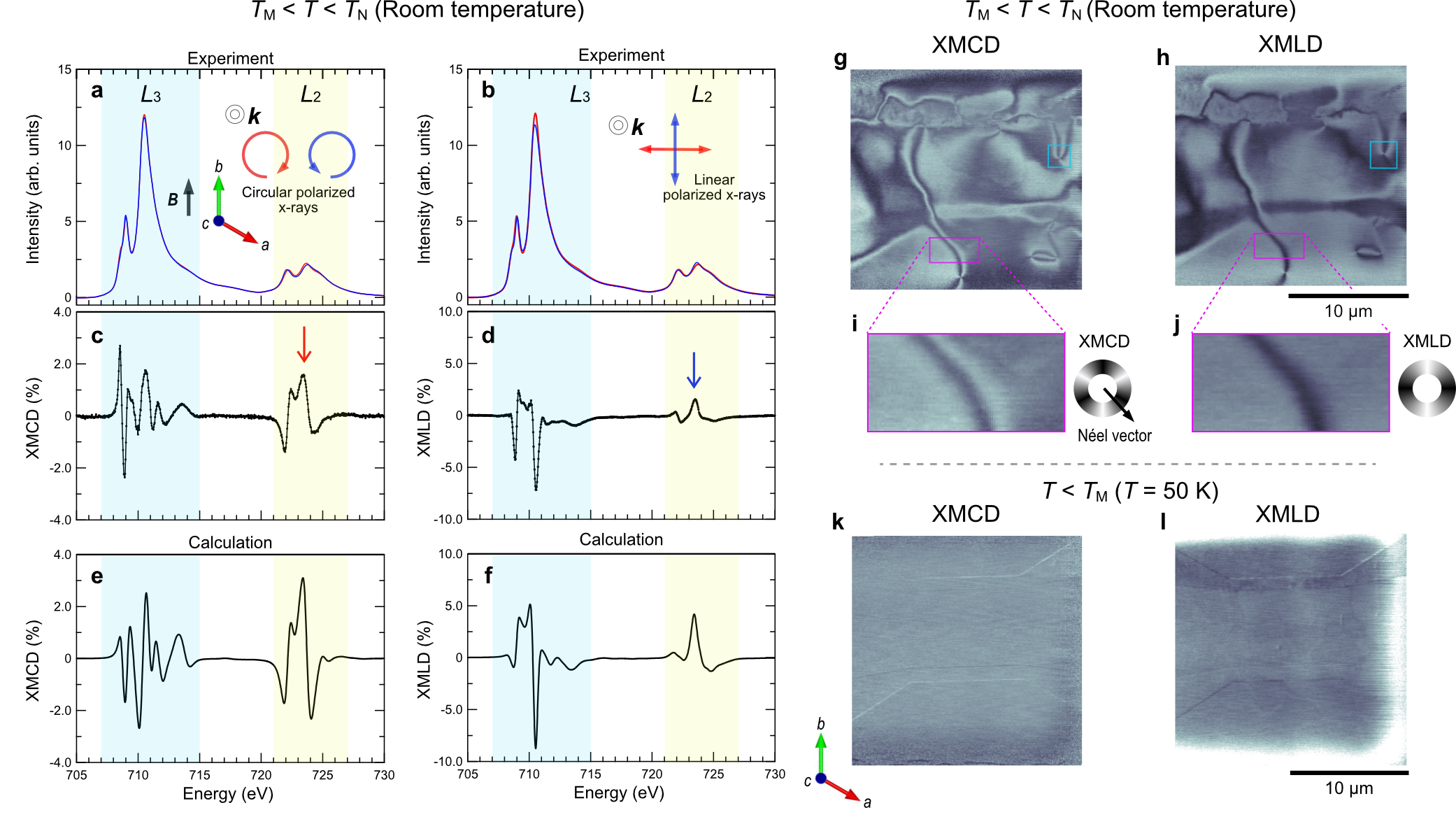}
\end{center}
\caption{X-ray magnetic circular and linear dichroism (XMCD and XMLD) of $\alpha$-Fe$_2$O$_3$ across Fe $L_{2,3}$-edges and corresponding calculations.
\textbf{a,b} X-ray absorption spectra acquired with circular right/linearly horizontal (red line) and circular left/linearly vertical (blue line) polarizations were measured by integral TEY measurements (Sample \#1) above $T_{\mathrm{M}}$. A magnetic field of approximately $100~\mathrm{mT}$ was applied within the $c$-plane.
\textbf{c,d} Corresponding experimental XMCD and XMLD spectra. 
\textbf{e,f} Calculated XMCD and XMLD spectra.
\textbf{g,h} STXM images of $\alpha$-Fe$_2$O$_3$ (Sample \#2) were recorded at the Fe $L_2$-edge, 724.25 eV for circular polarizations and 724.5 eV for linear polarization (nominal values) indicated by red and blue arrows in \textbf{c} and \textbf{d}. 
XMCD and XMLD images at room temperature show altermagnetic domains and N\'eel vector, respectively.  
\textbf{i},\textbf{j} Zoomed XMCD and XMLD images of the regions indicated by magenta boxes in \textbf{g} and \textbf{h} exhibit different contrast modulation, consistent with a 180$^\circ$ domain wall. Dependence of XMCD and XMLD contrasts on the in-plane $\textbf{L}$-orientation as indicated by the contrast wheels. At 50 K, all contrast vanished in the \textbf{k} XMCD and \textbf{l} XMLD images corresponding to \textbf{g} and \textbf{h}.}
\label{fig_2}
\end{figure*}

In this study, we demonstrate the presence of nanoscale spin textures with locally varying altermagnetic properties in $\alpha$-Fe$_2$O$_3$ (Hematite) using nanospectroscopic X-ray imaging. 
Long known as a naturally occurring antiferromagnet ($T_\mathrm{N}\sim950$~ K), $\alpha$-Fe$_2$O$_3$ has received recent attention due to its ability to host nanoscale topological textures~\cite{Chmiel18,Jani21,Tan24}, and more recently, due to its prediction as a $g$-wave altermagnetic candidate~\cite{Smejkal22a,Smejkal22b}, that has been confirmed with AHE~\cite{Galindez25}, XMCD~\cite{Galindez25}, and MOKE~\cite{Luo26,Pan26}. 
The magnetic structure is shown in Figure~\ref{fig_1}\textbf{a}, where the rhombohedral unit cell and $g$-wave spin-polarized electronic structure can be seen in Fig.~\ref{fig_1}\textbf{b}
[see the Supplementary Information (SI) for a more detailed discussion and symmetry analysis].
In $\alpha$-Fe$_2$O$_3$, there is a spin reorientation transition, known as the Morin transition, that occurs at $T_{\mathrm{M}} \sim 260$~K (Fig.~\ref{fig_1}$\bf{c}$)~\cite{Morin50}. Intriguingly, as one sweeps across this transition, the altermagnetic type and symmetry of the material will remain the same - however as the spins reorient, the emergent effects arising from SOC should disappear, as illustrated in Fig.~\ref{fig_1}$\bf{d,e}$.

Here, by probing spectroscopic X-ray magnetic circular dichroism, we observe finite XMCD at room temperature, which we confirm to be of altermagnetic origin with density functional theory (DFT) + dynamical mean-field theory (DMFT) calculations. We combine this altermagnetic XMCD with nanoscale imaging to identify the presence of a variety of altermagnetic spin textures.
By exploiting polarization-independent absorption contrast that distinguishes in-plane and out-of-plane $\bf{L}$-orientations, we identify the spin-reorientation associated with the Morin transition, and observe the disappearance of the XMCD--thus realizing the reorientation-induced \textit{on-off switching} of altermagnetic XMCD.
Despite the absence of XMCD, we characterize the altermagnetic domain configuration below $T_\mathrm{M}$ using this spectroscopic absorption contrast and identify the presence of nanoscale domain walls. Remarkably, these domain walls exhibit non-zero XMCD signal due to their local in-plane spin orientation, resulting in a finite altermagnetic response embedded in zero XMCD domains.  
Harnessing the $\bf{L}$-orientation dependent absorption contrast, together with XMLD and XMCD signals, we are able to identify winding spin textures at room temperature to be topological merons, again with an out-of-plane core that exhibits distinct emergent effects from the surrounding domains.
This modulation of emergent properties on the nanoscale in the form of domain walls and merons that exhibit distinct properties from the surrounding domains, demonstrates a key property of altermagnets that is particularly relevant for future information processing using altermagnetic spintronics.

\section{Dichroic altermagnetic spectroscopy}

In order to probe the altermagnetic nature of the material, we first perform spectroscopic X-ray magnetic circular dichroism measurements on a bulk single crystal of $\alpha$-Fe$_2$O$_3$ (Sample \#1). Details of the sample preparation, characterization, and measurements are provided in Methods and Supplementary information. 
We initially target the high temperature magnetic ordered state, where $\bL$ lies within the $c$-plane.
One of the challenges of performing non-spatially resolved spectroscopy is the prospect for multi-domain states that hinder the measurements~\cite{Hariki24}. In case of $\alpha$-Fe$_2$O$_3$, the Fe moments are slightly canted within this plane due to the Dzyaloshinskii--Moriya interaction~\cite{Dzyaloshinsky58,Moriya60}, resulting in a weak ferromagnetic moment ${\bf m}_{\rm FM}$. We exploit this weak ${\bf m}_{\rm FM}$ to induce a single $\bL$ domain state by applying an in-plane magnetic field during the measurements of the dichroic spectra~\cite{Suturin21}.
As schematically shown in the inset of Figure~\ref{fig_2}\textbf{a}, we apply an in-plane magnetic field of around 100~mT along a direction close to the $a$-axis to obtain a single-domain state for X-ray spectroscopy in total electron yield (TEY) mode at room temperature.

The X-ray absorption spectra (XAS) measured with left- and right-circularly polarized X-rays are shown in Fig.~\ref{fig_2}\textbf{a}, while the XAS measured with linearly polarized X-rays (horizontal and vertical) are shown in Fig,~\ref{fig_2}\textbf{b}.
In the XAS spectra, two distinct peaks are observed at both the Fe $L_{3}$ (approximately 707–715~eV) and $L_{2}$ edges (approximately 721–727~eV), consistent with previous studies~\cite{Kuiper93}. 
We then obtain the Fe $L_{2,3}$-edge XMCD spectrum (Figure~\ref{fig_2}\textbf{c}), and XMLD spectrum (Figure~\ref{fig_2}\textbf{d}) from the difference between left and right circular polarizations and between the two orthogonal linear polarizations, respectively.
Both spectra exhibit several sign reversals across the $L_3$ and $L_2$ edges, yet with clearly different spectral profiles.

To ascertain the origin of the observed dichroic signal, we performed DFT+DMFT calculations for the altermagnetic configuration. The calculated XMCD/XMLD spectra plotted in Fig.~\ref{fig_2}\textbf{e,f}, are consistent with the experimental data, confirming that the measured XMCD is indeed a signature of altermagnetism. 
Notably, the XMCD spectrum of $\alpha$-Fe$_2$O$_3$ closely resembles those reported for another altermagnet $\alpha$-MnTe~\cite{Hariki24, Amin24, Yamamoto25}. 
As XMCD can also originate from ferromagnetic moments, we further evaluated the possible contribution of a net ferromagnetic component.
In Supplementary Fig.~5, the calculated full optical conductivity tensor and the X-ray Hall vector at the Fe $L$ edge are provided, indicating that the contribution from a possible ferromagnetic moment is negligibly small compared with the intrinsic altermagnetic contribution. In addition, as shown in Supplementary Fig.~6, the XMCD spectrum calculated for a hypothetical ferromagnetic configuration with moments aligned along the $c$-axis exhibits a distinct line shape that differs significantly from the measured XMCD. 
We note that a similar calculation of the XMCD spectrum, with consistent results, has been presented in Ref.~\cite{Xie25,Ishii26}.

\section{Dichroic altmermagnetic imaging} 
We next explore the formation of nanoscale altermagnetic domains by scanning transmission X-ray microscopy (STXM) in the absence of magnetic fields on an x-ray transparent window (average thickness of ~150 nm) fabricated from a $\alpha$-Fe$_2$O$_3$ single crystal (Sample \#2, see Methods and SI for further details on sample fabrication and measurements). 
The sample geometry is shown in Extended Data Fig.~1. As shown there, lithographically patterned metallic Cu and Pt thin-film signal lines were fabricated on the sample for transport measurements; however, they were not used in the present work.
For the STXM imaging, the X-ray energy was tuned to values exhibiting sufficient photon transmission ($L_2$ edge) and strong dichroism, as identified from the spectroscopy (724.25 eV for XMCD and 724.5 eV for XMLD indicated by the arrows in Fig.~\ref{fig_2}\textbf{c},\textbf{d}).
As shown in the XMCD and XMLD images of Fig.~\ref{fig_2}\textbf{g} and \textbf{h}, a complex domain configuration was observed at room temperature.
First, at the metallic conductor regions indicated in Extended Data Fig.~1, the domain morphology is locally modified, implying a strain-induced effect.
Second, within this configuration, some domains of similar XMLD contrast exhibit opposite XMCD contrast.  
The characteristic features are highlighted by the zoomed images in Fig.~\ref{fig_2}\textbf{i},\textbf{j}, indicated by magenta boxes in Fig.~\ref{fig_2}\textbf{g},\textbf{h}. 
Between these two domains, a domain wall is observed, that exhibits a white-black-white-black contrast in XMCD, and a simpler white-black-white contrast in the XMLD image, consistent with the presence of an in-plane N\'eel-like  180$^\circ$ domain wall as observed in $\alpha$-MnTe~\cite{Yamamoto25}.
The time-reversal symmetry breaking and $\textbf{L}$ orientations resolved by XMCD and XMLD for the present experimental geometry are indicated by the color wheels in Fig~\ref{fig_2}, with further details provided in the Supplementary Figure~3.

Having established altermagnetic time-reversal symmetry breaking in the high-temperature phase, we move on to the low-temperature phase below the Morin transition, where $\bL$ reorients along the $c$-axis as illustrated in Figure~\ref{fig_1}\textbf{d}.
XMCD and XMLD images at $T_\mathrm{M}~>~T = 50$~K are presented in Fig.~\ref{fig_2}\textbf{k,l}.
The images are acquired under the same experimental conditions as those used for the high-temperature measurements shown in Fig.~\ref{fig_2}\textbf{g,h}.
Neither the XMCD nor the XMLD image exhibits significant dichroic contrast. This observation is consistent with the fact that XMLD is excluded by geometry and moreover with the prediction that because of relativistic effects, XMCD is not allowed by symmetry in this low-temperature phase/orientation, even though the underlying altermagnetic character of the system is expected to remain unchanged.

\begin{figure*}[!htbp]
\begin{center}
\includegraphics[width=2\columnwidth]{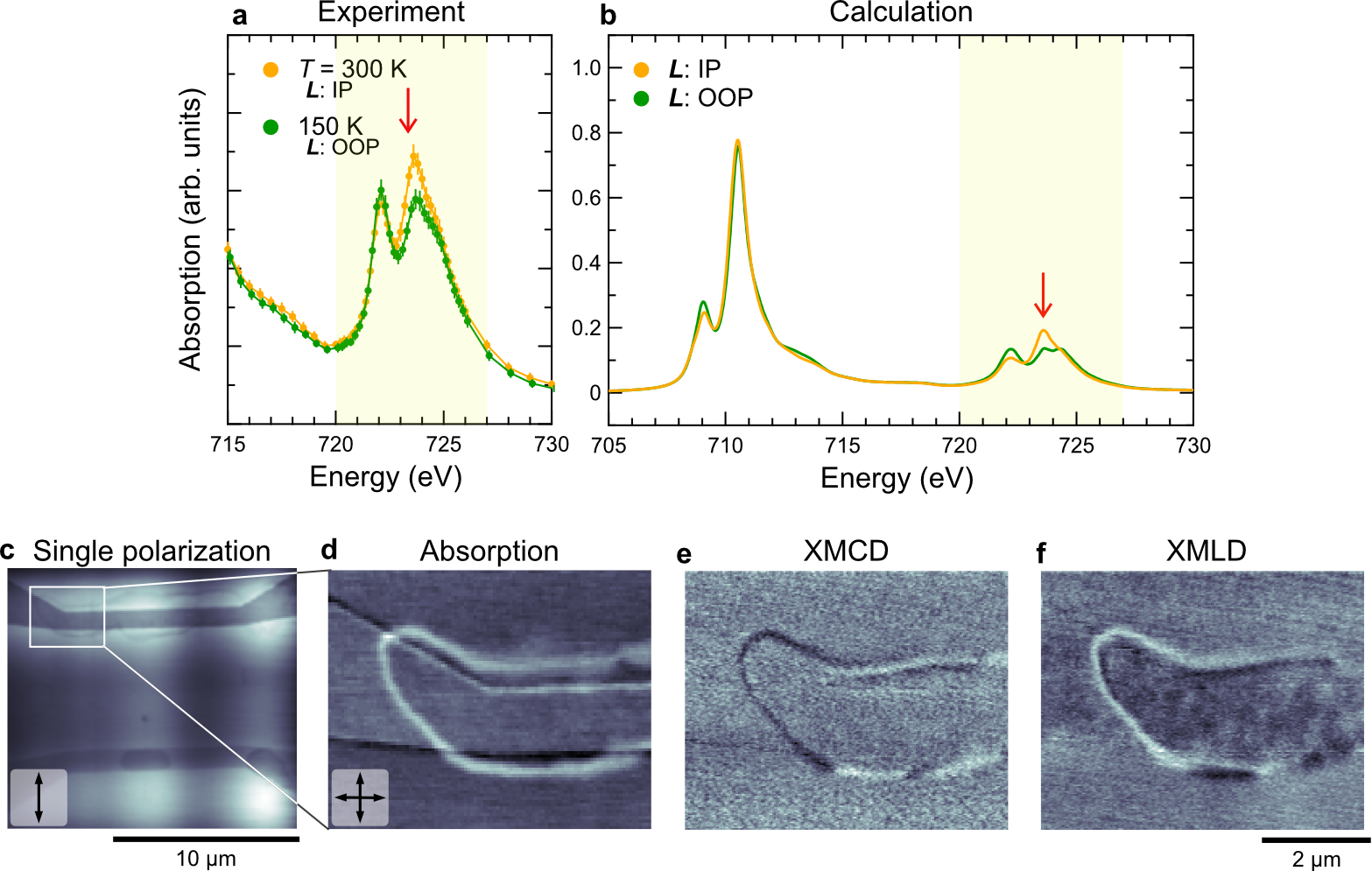}
\end{center}
\caption{Comparison between in-plane and out-of-plane $\bf{L}$-configurations. 
\textbf{a} Unpolarized experimental spectral data ($L_2$ edge) acquired from sum of linear polarization images (LH+LV) above and below $T_\mathrm{M}$ and 
\textbf{b} correspondingly calculated X-ray absorption spectra ($L_3$ and $L_2$ edges). 
\textbf{c} Overview image using a single polarization (LV) at 724.5 eV (indicated by red arrows in \textbf{a}) and corresponding magnified views of the region enclosed by the white box in \textbf{c}. The domain wall with in-plane $\bf{L}$-orientation is visible in \textbf{d} an absorption image obtained from the sum of linear horizontal and vertical polarization data. Here, the nonmagnetic absorption contrast was removed to emphasize the domain wall contrast (see  Supplementary Figure~4 for details). \textbf{e} XMCD image recorded at 724.25 eV and \textbf{f} XMLD image at 724.5 eV of the same region show dichroic signals only for the area covered by the in-plane domain wall.}
\label{fig_3}
\end{figure*}

\section{L-orientation and dichroic imaging below $T_M$}

To characterize this XMCD-bulk-forbidden low-temperature phase further, we turn, counterintuitively, to a
polarization-independent XAS effect that can be used to distinguish the $\mathbf{L}$-orientation (with respect to the crystal axes) in the high‐ and low‐temperature phases \cite{Kuiper93,Jani21} as illustrated in Figure~\ref{fig_1}\textbf{d}. Although the effect even occurs for arbitrary polarizations, we deliberately use unpolarized XAS spectra obtained by the sum of two orthogonal linear polarizations for avoiding regular dichroic effects in the following. 

As shown in Fig.~\ref{fig_3}\textbf{a}, we experimentally observed a clear change in the unpolarized XAS spectra extracted from absorption images acquired across $T_\mathrm{M}$.
These differences in the unpolarized XAS spectra are in good agreement with corresponding spectra calculated by DFT+DMFT calculations for in-plane and out-of-plane $\bL$ configurations, displayed in Fig.~\ref{fig_3}\textbf{b}.
It is worth pointing out that that the energy dependence and the magnitude of the contrast between the two $\mathbf{L}$-orientations cannot be anticipated from symmetry considerations alone, but arises from many-body interactions in the final state involving the core hole. 
By exploiting the difference in unpolarized x-ray absorption at certain energies between $(\mathbf{L}~||~c\mathrm{-axis})$ and $(\mathbf{L} \perp c\mathrm{-axis})$, it becomes possible to distinguish states of different $\mathbf{L}$-orientations in STXM imaging.

Following a reinitialization of the domain configuration by sweeping the sample temperature through the Morin transition, we observed an intriguing contrast near the upper metallic lead in the single-polarization image shown in Fig.~\ref{fig_3}\textbf{c}.
The image was acquired at an X-ray energy corresponding to the maximum $\mathbf{L}$-dependent absorption contrast (724.5~eV), at $T = 240$~K under an applied magnetic field of 250~mT along the hexagonal $a$-axis.
Fig.~\ref{fig_3}\textbf{d} shows an absorption image obtained by the sum of two linear polarizations zoomed into the region where this contrast appears.
In this figure, a white, curvilinear feature is clearly visible, with a full width at half maximum of $\sim160$~nm.
This bright contrast is consistent with a local reorientation of $\bf{L}$, and
corresponds to a domain wall with in-plane $\bf{L}$, that separates two out-of-plane $\mathcal{T}$-distinct altermagnetic domains.

Remarkably, when we perform XMCD (724.25~eV) and XMLD (724.5~eV) imaging of this curvilinear structure, shown in Fig.~\ref{fig_3}\textbf{e},\textbf{f}, a non-zero XMCD and XMLD contrast is observed in the region of the white line. This non-zero XMCD and XMLD is consistent with the in-plane orientation of the domain wall, 
which has the XMCD-allowed spin orientation typically associated with the high temperature phase, distinct from the surrounding domains in which XMCD is forbidden. 
The fact that XMCD is locally allowed only within the domain walls implies that other altermagnetic functionalities, such as the AHE, can be selectively activated at the nanoscale, highlighting these domain walls as promising functional elements for high-density, altermagnetic spintronic devices.

With this ability to characterize domain configurations in the altermagnetic phase for which observables such as XMCD and AHE are forbidden, a natural question is whether this contrast is unique to $\alpha$-Fe$_2$O$_3$, or whether it can be exploited in a wider range of $3d$ altermagnetic candidates. 
First, we note that the amplitude of the contrast is primarily governed by the interaction strength between the $3d$ electrons and the core hole, as reflected in the changes of the contrast across the multiplet peaks at the $L_2$ edge in $\alpha$-Fe$_2$O$_3$. 
In many localized $3d$ compounds, the $L_{2,3}$ edges are known to exhibit similar multiplet structures~\cite{groot90}, since this interaction strength does not vary significantly across the $3d$ elements, and therefore such $\bf{L}$-dependent XAS contrast can be expected for a wide range of systems. 
Second, the energy dependence of the contrast, and the energy at which it is maximized, can be predicted using simple ionic multiplet calculations~\cite{groot90}, as demonstrated in the Supplementary information - meaning that experimental access to both phases is not necessarily needed to be able to predict the nature of such an effect. As a result, we believe this unpolarized XAS contrast offers a robust and widely implementable route to characterize altermagnetic configurations, even when the primary domains do not exhibit any XMCD.

\section{Topological Altermagnetic Textures}

\begin{figure}[h]
\begin{center}
\includegraphics[width=1\columnwidth]{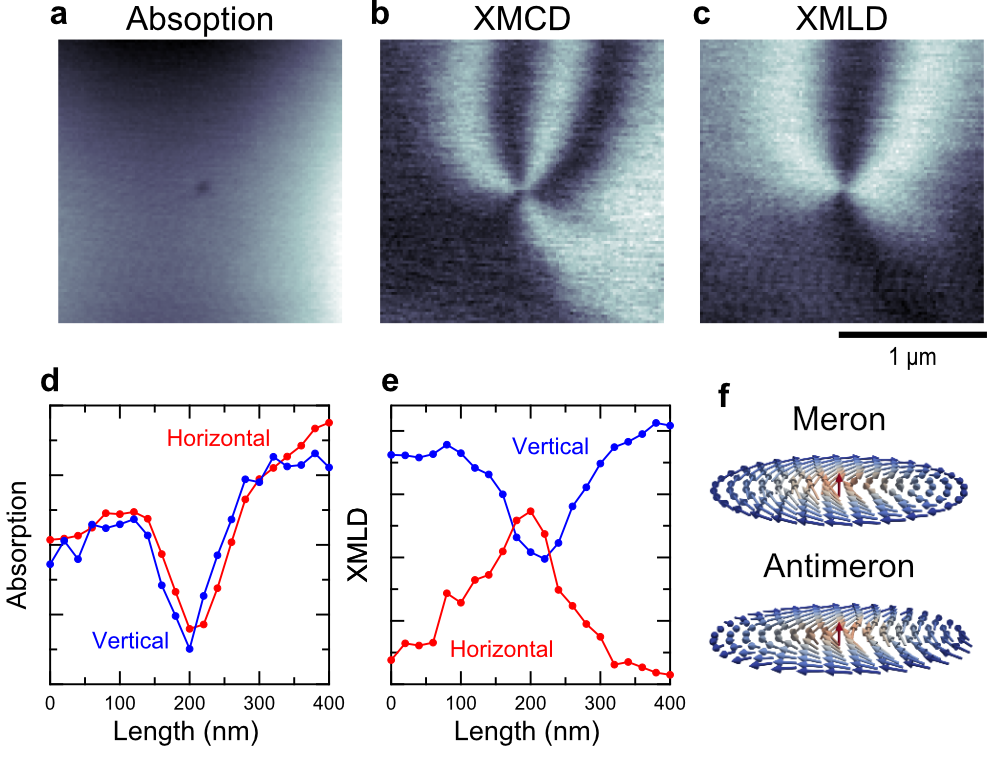}
\end{center}
\caption{Altermagnetic nanotexture at the region indicated by the cyan box in Fig. \ref{fig_2}\textbf{g} and \textbf{h}.  \textbf{a} Unpolarized absorption, \textbf{b} XMCD, and \textbf{c} XMLD images at nominally 724.25, 724.25 and 724.5 eV, respectively. The dark point in the absorption image represents a core with  out-of-plane $\bf{L}$-orientation. XMCD and XMLD images show dichroic signal due to $\mathcal{T}$-symmetry broken domains and the in-plane $\bf{L}$ orientations surrounding the core, respectively. 
The core shows a spatial extent of approximately $140\pm20$~nm from the line profile in \textbf{d} (extracted from \textbf{a}), and broad peaks across the core in both the absorption \textbf{d} and XMLD profiles \textbf{e} (extracted from \textbf{c}) indicate a continuous N\'eel vector rotation from out-of-plane to in-plane $\bf{L}$ orientations, consistent with the presence of a topological texture with $Q = \pm 1/2$ as schematically illustrated in \textbf{f}.}
\label{fig_4}
\end{figure}

Having identified the presence of an in-plane domain wall in the low temperature regime, we next explore whether similar textures exhibiting a mixture of relativistically distinct $\textbf{L}$-orientations can be observed in the high temperature regime. In particular, when we perform absorption contrast imaging at the region of the cyan box in Fig~\ref{fig_2}\textbf{g} or \textbf{h}, a distinct nanoscale dark dot is observed in the XAS image in Figure~\ref{fig_4}\textbf{a}, that is located at the center of rotationally-symmetric XMCD and XMLD patterns.  Indeed, as displayed in Fig.~\ref{fig_4}\textbf{b}, \textbf{c}, the XMCD and XMLD contrast is consistent with an in-plane winding texture such as a vortex or antivortex, as has been observed previously in $\alpha$-MnTe~\cite{Amin24,Yamamoto25}. However, the dark absorption contrast at the center of the structure indicates that at the core of the vortex, $\bL$ tilts out-of-plane, analogous to what is seen in ferromagnets.
Thus, it forms a meron-type topological texture with a skyrmion number of Q=±1/2, characterized by a three-dimensionally winding spin configuration.
Indeed, when we take line profiles of the unpolarized XAS and XMLD contrasts through the core along the horizontal and vertical spatial directions (Extended Data Fig.~2), one can see that the XMLD line profiles show their maximum and minimum along the horizontal and vertical directions at the same location where there is a maximum in the $\mathbf{L}$-dependent absorption contrast, as shown in Fig.~\ref{fig_4}\textbf{d}, \textbf{e}.
These peaks have a full width half maximum of $140\pm20$~nm, confirming that the width of the core is significantly larger than our spatial resolution, and thus that the observed vanishing of XMLD/XMCD signal does not occur due to a spatial resolution artefact.
This reduction in dichroism, coinciding with the $\bL$-dependent absorption contrast, confirms that $\bf{L}$ tilts out-of-plane in the core, and indicates that the region in the core is in a relativistically distinct altermagnetic state from the surrounding in-plane configuration. 

The confirmation of nanoscale complex spin textures--a domain wall in the low temperature regime, and a meron in the high temperature regime--that consist of a mixture of two relativistically-distinct altermagnetic states is not only of interest from a fundamental point of view.  
Yet the prospect of realizing stable, localized variations in emergent properties, such as XMCD or AHE, whose magnitude and sign are governed by the local orientation of the N\'eel vector and can be controlled by magnetic fields or electric currents, represents an important step towards developing altermagnetic spintronics.

\section{Conclusions}
In conclusion, we have harnessed X-ray spectroscopy to characterize the altermagnetism in $\alpha$‐Fe$_2$O$_3$. 
Using XMCD spectroscopy, we confirm the altermagnetic nature of the material and, by exploiting the temperature-induced reorientation of the N\'eel vector, identify two distinct spin–orbit-coupling-induced altermagnetic phases above and below the Morin transition. These results confirm that $\alpha$‐Fe$_2$O$_3$ exhibits different emergent properties dependent on its N\'eel vector orientation despite being the same type of $g$-wave altermagnet.
Moreover, by identifying an $\mathbf{L}$-dependent X-ray absorption contrast (independent of the X-ray polarization), we were able to gain insight into the low temperature phase of the material for which XMCD is not allowed, and further identify the presence of nanoscale spin textures including domain walls and merons.
The presence of these nanoscale textures composed with different N\'eel vector orientation is particularly relevant in the context of potential applications. As the same symmetries lead to both XMCD and AHE, the identification of a domain wall with non-zero XMCD embedded between zero-XMCD domains establishes the possibility of isolated nanotextures that alone produce spintronic signals, offering a route towards high-density altermagnetic devices. 
The underlying on-and-off switching of emergent properties is a general consequence of the N\'eel vector orientation in altermagnets, and hence such local modulations are expected to be present in various systems with similar magnetic symmetries, paving the way for nanoscale device applications of a broad class of altermagnetic materials.


\subsection{Methods}

\subsubsection{Sample preparation}
All samples were fabricated from single-crystalline $\alpha$-Fe$_2$O$_3$ naturally grown in the Kalahari region of Botswana and supplied by Surfacenet GmbH (Germany).
The samples were shaped to sizes of (3 x 3 x 0.1) mm$^{3}$ (Sample \#1, TEY) and (5 x 5 x 0.1) mm$^{3}$ (Sample \#2, STXM) by the supplier, with the $c$-axis being perpendicular to the sample surface. The supplier further polished the samples from one side (Sample \#1) and both sides (Sample \#2), respectively. 

Sample \#2 was thinned down to soft x-ray transparent levels ($\sim$$150$~nm) via plasma focused ion beam (PFIB)lithography using a Xe plasma. The processing follows the protocol outlined in \cite{Mayr21}. An acceleration voltage of 30~kV was used for the etching, and the current was varied between 2.5~uA at the start of the etching and 4~nA upon reaching the target thickness. The end point detection for the etching procedure was determined by observing secondary electron emission from the “far side” surface of the $\alpha$-Fe$_2$O$_3$ membrane with scanning electron microscopy at 10~kV extractor voltage. To protect the etched membrane, a 5~nm thick carbon film was deposited from the back side of the $\alpha$-Fe$_2$O$_3$ sample after the etching by means of thermal evaporation using a CCU-010 HV evaporator from Safematic GmbH.
Metallic thin film copper conductors for other experimental purposes (transport and dynamic excitation) were fabricated onto these windows by means of electron beam lithography, thin film deposition, and lift-off processing. 

\subsubsection{Magnetometry}
The magnetization of Sample \#1 was measured in the temperature range from 200 to 300 K using a commercial superconducting quantum interference device (SQUID) magnetometer in vibrating‐sample mode (VSM) (MPMS, Quantum Design). A magnetic field of 0.01 T was applied along the $m$ axis. As shown in Fig.~\ref{fig_1}\textbf{d}, a clear drop in the magnetization was observed at $T_\mathrm{onset} = 257.9$ K, corresponding to the Morin transition~\cite{Morin50}.

\subsubsection{X-Ray Absorption Spectroscopy}
X-ray absorption spectroscopy (XAS) of Sample \#1 was carried out at the XUV diffractometer of the UE46-PGM1 undulator beamline with circularly and linearly polarized x-rays by measuring the total electron yield via sample drain current. The photon energy resolution was set to 90 meV at the Fe-$L_{2,3}$ resonances. 
The circular polarization at the 3rd harmonic is not 100\% but rather approximately 90\%.
For measurements with applied field, the sample was placed between two SmCo permanent magnets at a distance
of 4~mm. The field in the center of the sample was approx. 100~mT.

The x-ray magnetic circular and linear dichroism spectra measured were processed in the following way. 
The measured intensity of x-rays $I$ is calibrated using the incident photon flux $I_{0}$.
The linear background is removed from the spectra ($I/I_{0}$). 
The intensity is normalized by the difference between $L_{3}$ pre-edge and $L_{2}$ post-edge. 
The dichroic spectra were then calculated in the following way.
\begin{equation}
\mathrm{XMCD/XMLD} = 100 \times \frac{(I_{\mathrm{norm}}^{-/\mathrm{LV}} - I_{\mathrm{norm}}^{+/\mathrm{LH}})} {A_{\mathrm{max}}} \%.
\label{eq:eq_1}
\end{equation}
Here, \(I_{\mathrm{norm}}^{+}\) and \(I_{\mathrm{norm}}^{-}\) correspond to the normalized absorption spectra measured using right- and left-circularly polarized X-rays, respectively, and are used for the XMCD calculation. 
Similarly, \(I_{\mathrm{norm}}^{\mathrm{LH}}\) and \(I_{\mathrm{norm}}^{\mathrm{LV}}\) denote the normalized absorption spectra measured using linearly polarized X-rays with horizontal (LH) and vertical (LV) polarization, respectively, and are used for the XMLD calculation. 
The difference of opposite polarization is divided by $A_{\mathrm{max}}$ which is the maximum absorption of the material at the $L_{3}$ edge. Calculating the XMCD/XMLD as a percentage of this value allows us to obtain thickness independent values. These values can then be quantitatively compared to different energies, and furthermore  with predicted XMCD/XMLD values from the LDA+DMFT AIM calculations.

\subsubsection{Scanning Transmission X-Ray Microscopy}
Sample \#2 was imaged using scanning transmission x-ray microscopy (STXM) at at the MAXYMUS \cite{Weigand22} end station of the UE46 beamline of the BESSY II synchrotron radiation facility of Helmholtz-Zentrum Berlin. X-rays with defined polarization (linear horizontal, linear vertical, right circular, or left circular) and at energies corresponding to the Fe-$L_{2,3}$ absorption egdes are provided by an undulator insertion device, operating in the third harmonics. The photon energy is further confined in the beamline to $\sim$50...100 meV by a subsequent monochromator. For STXM imaging, the X-rays are focused onto the sample by a diffractive Fresnel zone plate to a spot of $\sim$25 nm diameter and both, undiffracted light and higher diffraction orders are blocked by an order selective aperture (OSA) in front of the sample. Photons transmitted through the sample are detected by a point detector. Raster scanning the sample through the beam yields a two-dimensional image. Images were collected at different polarizations and energies for achieving nanoscopic spectral dichroism. 
The transmission images recorded with right-/left-circular and horizontal-/vertical-linear polarized x-rays are denoted as 
\( I^{\mathrm{+/LH}} \) and \( I^{\mathrm{-/LV}} \), respectively.
In order to correct for differences in incident flux or beamline intensity drift between the two polarization states, a normalization coefficient
\begin{equation}
a = 
\frac{\langle I^{\mathrm{-}} \rangle}{\langle I^{\mathrm{+}} \rangle}
\label{eq:eq_2}
\end{equation}
was applied, where \( \langle \cdot \rangle \) represents the spatial average over the image.
The image with right-circular or horizontal-linear polarized x-ray was normalized as
\begin{equation}
I^{\mathrm{+/LH}}_{\mathrm{norm}} = a \, I^{\mathrm{+/LH}} .
\label{eq:eq_3}
\end{equation}
This procedure ensures that the average transmission level is consistent between the two polarization states.
The dichroic image in absorption geometry was then computed from the normalized transmission intensities as
\begin{equation}
\mathrm{XMCD/XMLD} = 
\frac{-\ln I^{\mathrm{-/LV}} + \ln I^{\mathrm{+/LH}}_{\mathrm{norm}}}{2}.
\label{eq:eq_4}
\end{equation}
In addition to the dichroic signal, the polarization-averaged intensity
\begin{equation}
I_{\mathrm{avg}} = \frac{I^{\mathrm{-/LV}} + I^{\mathrm{+/LH}}_{\mathrm{norm}}}{2}
\label{eq:eq_5}
\end{equation}
and the absorption contrast
\begin{equation}
\mathrm{Absorption} = 
\frac{-(\ln I^{\mathrm{-/LV}} + \ln I^{\mathrm{+/LH}}_{\mathrm{norm}})}{2}
\label{eq:eq_6}
\end{equation}
were calculated.

\subsubsection{X-Ray Magnetic Dichroism Calculations}

We simulate the Fe $L$-edge spectra shown in Fig.~\ref{fig_2} and Fig.~\ref{fig_3} using density functional theory combined with dynamical mean-field theory (DFT+DMFT), following the implementation of Refs.~\cite{Hariki18,Hariki20,Li23}. First, we performed a standard DFT+DMFT calculation in the magnetically ordered phase for the experimental crystal structure of $\alpha$-Fe$_2$O$_3$, as described in Ref.~\cite{Li23}. The Fe $L$-edge spectral intensities were then computed using an Anderson impurity model (AIM), incorporating the spin-dependent hybridization densities $\Delta(\varepsilon)$ obtained from DFT+DMFT, which represent the Weiss field in the altermagnetic ordered state. The same parameters for the 3$d$–3$d$ and 2$p$–3$d$ interactions, as well as the charge-transfer energy $\Delta_{\rm CT}$, were adopted as in the previous study on Fe $L$-edge resonant inelastic x-ray scattering of $\alpha$-Fe$_2$O$_3$~\cite{Li23}. Since the core hole created by the x-rays is fully localized at the excited Fe site, the total $L$-edge spectrum is obtained by summing the contributions from the AIMs corresponding to the four Fe sites in the unit cell, as successfully applied in previous works for $\alpha$-MnTe~\cite{Hariki24}.

\subsection{Acknowledgments}
We thank Ale\v{s} Hrabec for the Pt deposition during the sample fabrication as well as Benny B\"ohm and Fabian Ganss for initial SQUID measurements on similar samples.
Part of the sample fabrication was carried out at the cleanroom facility of the Paul Scherrer Institute.
We acknowledge the Helmholtz-Zentrum Berlin f\"ur Materialien und Energie for the allocation of synchrotron radiation beamtime.
S.M. was funded by the Swiss National Science Foundation (Grant Agreement No. 172517).
A.H. was supported by JSPS KAKENHI Grant Numbers 25K00961, 25K07211, the 2025 Osaka Metropolitan University (OMU) Strategic Research Promotion Project (Young Researcher). J.K. was supported by the project Quantum materials for applications in sustainable technologies (QM4ST), funded as project No. CZ.02.01.01/00/22 008/0004572 by Programme Johannes Amos Commenius, call Excellent Research, and by the Ministry of Education, Youth and Sports of the Czech Republic through the e-INFRA CZ (ID:90254).
L.\v{S}. acknowledges funding from the ERC Starting Grant No. 101165122 and Deutsche Forschungs-gemeinschaft (DFG) grant no. TRR 288-7422213477 (Projects A09 and B05).

\bibliography{fe2o3}

@article{Hariki24,
  title = {{X-Ray Magnetic Circular Dichroism in Altermagnetic $\alpha$-{MnTe}}},
  author = {Hariki, A. and Dal Din, A. and Amin, O. J. and Yamaguchi, T. and Badura, A. and Kriegner, D. and Edmonds, K. W. and Campion, R. P. and Wadley, P. and Backes, D. and Veiga, L. S. I. and Dhesi, S. S. and Springholz, G. and \ifmmode \check{S}\else \v{S}\fi{}mejkal, L. and V\'yborn\'y, K. and Jungwirth, T. and Kune\ifmmode \check{s}\else \v{s}\fi{}, J.},
  journal = {Phys. Rev. Lett.},
  volume = {132},
  issue = {17},
  pages = {176701},
  numpages = {7},
  year = {2024},
  month = {Apr},
  publisher = {American Physical Society},
  doi = {10.1103/PhysRevLett.132.176701},
  url = {https://link.aps.org/doi/10.1103/PhysRevLett.132.176701}
}

@article{Groot90,
  title = {2p x-ray absorption of 3d transition-metal compounds: An atomic multiplet description including the crystal field},
  author = {de Groot, F. M. F. and Fuggle, J. C. and Thole, B. T. and Sawatzky, G. A.},
  journal = {Phys. Rev. B},
  volume = {42},
  issue = {9},
  pages = {5459--5468},
  numpages = {0},
  year = {1990},
  month = {Sep},
  publisher = {American Physical Society},
  doi = {10.1103/PhysRevB.42.5459},
  url = {https://link.aps.org/doi/10.1103/PhysRevB.42.5459}
}

@article{Hariki20,
  title = {$\mathrm{LDA}+\mathrm{DMFT}$ approach to resonant inelastic x-ray scattering in correlated materials},
  author = {Hariki, Atsushi and Winder, Mathias and Uozumi, Takayuki and Kune\ifmmode \check{s}\else \v{s}\fi{}, Jan},
  journal = {Phys. Rev. B},
  volume = {101},
  issue = {11},
  pages = {115130},
  numpages = {9},
  year = {2020},
  month = {Mar},
  publisher = {American Physical Society},
  doi = {10.1103/PhysRevB.101.115130},
  url = {https://link.aps.org/doi/10.1103/PhysRevB.101.115130}
}

@article{Jaeschke25,
  title={Atomic altermagnetism},
  author={Jaeschke-Ubiergo, Rodrigo and Bharadwaj, Venkata-Krishna and Campos, Warlley and Zarzuela, Ricardo and Biniskos, Nikolaos and Fernandes, Rafael M and Jungwirth, Tomas and Sinova, Jairo and {\v{S}}mejkal, Libor},
  journal={arXiv preprint arXiv:2503.10797},
  year={2025}
}

@article{Hariki18,
  title = {{Continuum Charge Excitations in High-Valence Transition-Metal Oxides Revealed by Resonant Inelastic X-Ray Scattering}},
  author = {Hariki, Atsushi and Winder, Mathias and Kune\ifmmode \check{s}\else \v{s}\fi{}, Jan},
  journal = {Phys. Rev. Lett.},
  volume = {121},
  issue = {12},
  pages = {126403},
  numpages = {6},
  year = {2018},
  month = {Sep},
  publisher = {American Physical Society},
  doi = {10.1103/PhysRevLett.121.126403},
  url = {https://link.aps.org/doi/10.1103/PhysRevLett.121.126403}
}

@article{Li23,
  title = {{Single- and Multimagnon Dynamics in Antiferromagnetic $\alpha$-{Fe}$_{2}${O}$_{3}$ Thin Films}},
  author = {Li, Jiemin and Gu, Yanhong and Takahashi, Yoshihiro and Higashi, Keisuke and Kim, Taehun and Cheng, Yang and Yang, Fengyuan and Kune\ifmmode \check{s}\else \v{s}\fi{}, Jan and Pelliciari, Jonathan and Hariki, Atsushi and Bisogni, Valentina},
  journal = {Phys. Rev. X},
  volume = {13},
  issue = {1},
  pages = {011012},
  numpages = {9},
  year = {2023},
  month = {Feb},
  publisher = {American Physical Society},
  doi = {10.1103/PhysRevX.13.011012},
  url = {https://link.aps.org/doi/10.1103/PhysRevX.13.011012}
}

@article{Yamamoto25,
  title = {{Altermagnetic nanotextures revealed in bulk {MnTe}}},
  author = {Yamamoto, Rikako and Turnbull, Luke Alexander and Schmidt, Marcus and Corsaletti Filho, Jos\'e Claudio and Binger, Hayden Jeffrey and Di Pietro Mart\'{\i}nez, Marisel and Weigand, Markus and Finizio, Simone and Prots, Yurii and Ferguson, George Matthew and Vool, Uri and Wintz, Sebastian and Donnelly, Claire},
  journal = {Phys. Rev. Appl.},
  volume = {24},
  issue = {3},
  pages = {034037},
  numpages = {11},
  year = {2025},
  month = {Sep},
  publisher = {American Physical Society},
  doi = {10.1103/dp7v-qszq},
  url = {https://link.aps.org/doi/10.1103/dp7v-qszq}
}

@article{Galindez25,
author = {Galindez-Ruales, Edgar and Gonzalez-Hernandez, Rafael and Schmitt, Christin and Das, Shubhankar and Fuhrmann, Felix and Ross, Andrew and Golias, Evangelos and Akashdeep, Akashdeep and Lünenbürger, Laura and Baek, Eunchong and Yang, Wanting and Šmejkal, Libor and Krishna, Venkata and Jaeschke-Ubiergo, Rodrigo and Sinova, Jairo and Rothschild, Avner and You, Chun-Yeol and Jakob, Gerhard and Kläui, Mathias},
title = {{Revealing the Altermagnetism in Hematite via {XMCD} Imaging and Anomalous Hall Electrical Transport}},
journal = {Advanced Materials},
volume = {n/a},
number = {n/a},
pages = {e05019},
year = {2025},
doi = {https://doi.org/10.1002/adma.202505019},
url = {https://advanced.onlinelibrary.wiley.com/doi/abs/10.1002/adma.202505019}
}

@article{Bhowal24,
  title={Ferroically ordered magnetic octupoles in d-wave altermagnets},
  author={Bhowal, Sayantika and Spaldin, Nicola A},
  journal={Physical Review X},
  volume={14},
  number={1},
  pages={011019},
  year={2024},
  publisher={APS}
}

@article{Smejkal20,
  title={Crystal time-reversal symmetry breaking and spontaneous {H}all effect in collinear antiferromagnets},
  author={{\v{S}}mejkal, Libor and Gonz{\'a}lez-Hern{\'a}ndez, Rafael and Jungwirth, Tom{\'a}{\v{s}} and Sinova, Jairo},
  journal={Science advances},
  volume={6},
  number={23},
  pages={eaaz8809},
  year={2020},
  publisher={American Association for the Advancement of Science}
}

@article{GonzalezBetancourt23,
  title = {{Spontaneous Anomalous Hall Effect Arising from an Unconventional Compensated Magnetic Phase in a Semiconductor}},
  author = {Gonzalez Betancourt, R. D. and Zub\'a\ifmmode \check{c}\else \v{c}\fi{}, J. and Gonzalez-Hernandez, R. and Geishendorf, K. and \ifmmode \check{S}\else \v{S}\fi{}ob\'a\ifmmode \check{n}\else \v{n}\fi{}, Z. and Springholz, G. and Olejn\'{\i}k, K. and \ifmmode \check{S}\else \v{S}\fi{}mejkal, L. and Sinova, J. and Jungwirth, T. and Goennenwein, S. T. B. and Thomas, A. and Reichlov\'a, H. and \ifmmode \check{Z}\else \v{Z}\fi{}elezn\'y, J. and Kriegner, D.},
  journal = {Phys. Rev. Lett.},
  volume = {130},
  issue = {3},
  pages = {036702},
  numpages = {7},
  year = {2023},
  month = {Jan},
  publisher = {American Physical Society},
  doi = {10.1103/PhysRevLett.130.036702},
  url = {https://link.aps.org/doi/10.1103/PhysRevLett.130.036702}
}

@article{Verbeek24,
  title={Nonrelativistic ferromagnetotriakontadipolar order and spin splitting in hematite},
  author={Verbeek, XH and Voderholzer, David and Sch{\"a}ren, Stefan and Gachnang, Yannick and Spaldin, NA and Bhowal, Sayantika},
  journal={Physical Review Research},
  volume={6},
  number={4},
  pages={043157},
  year={2024},
  publisher={APS}
}

@Article{Amin24,
author={Amin, O. J.
and Dal Din, A.
and Golias, E.
and Niu, Y.
and Zakharov, A.
and Fromage, S. C.
and Fields, C. J. B.
and Heywood, S. L.
and Cousins, R. B.
and Maccherozzi, F.
and Krempask{\'y}, J.
and Dil, J. H.
and Kriegner, D.
and Kiraly, B.
and Campion, R. P.
and Rushforth, A. W.
and Edmonds, K. W.
and Dhesi, S. S.
and {\v{S}}mejkal, L.
and Jungwirth, T.
and Wadley, P.},
title={Nanoscale imaging and control of altermagnetism in {MnTe}},
journal={Nature},
year={2024},
month={Dec},
day={01},
volume={636},
number={8042},
pages={348-353},
issn={1476-4687},
doi={10.1038/s41586-024-08234-x},
url={https://doi.org/10.1038/s41586-024-08234-x}
}

@article{Reimers24,
  title={Direct observation of altermagnetic band splitting in {CrSb} thin films},
  author={Reimers, Sonka and Odenbreit, Lukas and {\v{S}}mejkal, Libor and Strocov, Vladimir N and Constantinou, Procopios and Hellenes, Anna B and Jaeschke Ubiergo, Rodrigo and Campos, Warlley H and Bharadwaj, Venkata K and Chakraborty, Atasi and Denneulin, Thibaud and Shi, Wen and Dunin-Borkowski, Rafal E. and Das, Suvadip and Kl\"aui, Mathias and Sinova, Jairo and Jourdan, Martin},
  journal={Nat. Commun.},
  volume={15},
  number={1},
  pages={2116},
  year={2024},
  publisher={Nature Publishing Group UK London}
}

@article{Ding24,
  title = {{Large Band Splitting in $g$-Wave Altermagnet CrSb}},
  author = {Ding, Jianyang and Jiang, Zhicheng and Chen, Xiuhua and Tao, Zicheng and Liu, Zhengtai and Li, Tongrui and Liu, Jishan and Sun, Jianping and Cheng, Jinguang and Liu, Jiayu and Yang, Yichen and Zhang, Runfeng and Deng, Liwei and Jing, Wenchuan and Huang, Yu and Shi, Yuming and Ye, Mao and Qiao, Shan and Wang, Yilin and Guo, Yanfeng and Feng, Donglai and Shen, Dawei},
  journal = {Phys. Rev. Lett.},
  volume = {133},
  issue = {20},
  pages = {206401},
  numpages = {7},
  year = {2024},
  month = {Nov},
  publisher = {American Physical Society},
}

@article{Hoyer25,
  title = {Altermagnetic splitting of magnons in hematite $\alpha$-{Fe}$_{2}${O}$_{3}$},
  author = {Hoyer, Rhea and Stavropoulos, P. Peter and Razpopov, Aleksandar and Valentí, Roser and {\v{S}}mejkal, Libor and Mook, Alexander},
  journal = {Phys. Rev. B},
  volume = {112},
  issue = {6},
  pages = {064425},
  numpages = {22},
  year = {2025},
  month = {Aug},
  publisher = {American Physical Society},
  doi = {10.1103/fgc1-5blp},
  url = {https://link.aps.org/doi/10.1103/fgc1-5blp}
}

@article{Chmiel18,
	title = {Observation of magnetic vortex pairs at room temperature in a planar $\alpha$-{Fe}$_{2}${O}$_{3}$/{Co} heterostructure},
	volume = {17},
	issn = {1476-1122, 1476-4660},
	url = {https://www.nature.com/articles/s41563-018-0101-x},
	doi = {10.1038/s41563-018-0101-x},
	number = {7},
	urldate = {2025-10-14},
	journal = {Nature Materials},
	author = {Chmiel, F. P. and Waterfield Price, N. and Johnson, R. D. and Lamirand, A. D. and Schad, J. and Van Der Laan, G. and Harris, D. T. and Irwin, J. and Rzchowski, M. S. and Eom, C.-B. and Radaelli, P. G.},
	month = {Jul},
	year = {2018},
	pages = {581--585},
}

@article{Lee24,
  title={Broken {K}ramers degeneracy in altermagnetic {MnTe}},
  author={Lee, Suyoung and Lee, Sangjae and Jung, Saegyeol and Jung, Jiwon and Kim, Donghan and Lee, Yeonjae and Seok, Byeongjun and Kim, Jaeyoung and Park, Byeong Gyu and {\v{S}}mejkal, Libor and others},
  journal={Physical review letters},
  volume={132},
  number={3},
  pages={036702},
  year={2024},
  publisher={APS}
}

@article{Krempasky24,
	title = {Altermagnetic lifting of {K}ramers spin degeneracy},
	volume = {626},
	issn = {0028-0836, 1476-4687},
	url = {https://www.nature.com/articles/s41586-023-06907-7},
	doi = {10.1038/s41586-023-06907-7},
	number = {7999},
	urldate = {2025-05-28},
	journal = {Nature},
	author = {Krempaský, J. and Šmejkal, L. and D’Souza, S. W. and Hajlaoui, M. and Springholz, G. and Uhlířová, K. and Alarab, F. and Constantinou, P. C. and Strocov, V. and Usanov, D. and Pudelko, W. R. and González-Hernández, R. and Birk Hellenes, A. and Jansa, Z. and Reichlová, H. and Šobáň, Z. and Gonzalez Betancourt, R. D. and Wadley, P. and Sinova, J. and Kriegner, D. and Minár, J. and Dil, J. H. and Jungwirth, T.},
	month = feb,
	year = {2024},
	pages = {517--522},
}

@article{Osumi24,
	title = {Observation of a giant band splitting in altermagnetic {MnTe}},
	volume = {109},
	issn = {2469-9950, 2469-9969},
	url = {https://link.aps.org/doi/10.1103/PhysRevB.109.115102},
	doi = {10.1103/PhysRevB.109.115102},
	number = {11},
	urldate = {2025-10-14},
	journal = {Physical Review B},
	author = {Osumi, T. and Souma, S. and Aoyama, T. and Yamauchi, K. and Honma, A. and Nakayama, K. and Takahashi, T. and Ohgushi, K. and Sato, T.},
	month = mar,
	year = {2024},
	pages = {115102},
}

@article{Smejkal22a,
  title = {{Beyond Conventional Ferromagnetism and Antiferromagnetism: A Phase with Nonrelativistic Spin and Crystal Rotation Symmetry}},
  author = {\ifmmode \check{S}\else \v{S}\fi{}mejkal, Libor and Sinova, Jairo and Jungwirth, Tomas},
  journal = {Phys. Rev. X},
  volume = {12},
  issue = {3},
  pages = {031042},
  numpages = {16},
  year = {2022},
  month = {Sep},
  publisher = {American Physical Society},
  doi = {10.1103/PhysRevX.12.031042},
  url = {https://link.aps.org/doi/10.1103/PhysRevX.12.031042}
}

@article{Smejkal22b,
  title={Emerging research landscape of altermagnetism},
  author={{\v{S}}mejkal, Libor and Sinova, Jairo and Jungwirth, Tomas},
  journal={Physical Review X},
  volume={12},
  number={4},
  pages={040501},
  year={2022},
  publisher = {American Physical Society},
}

@article{Suturin21,
	title = {X-ray magnetic linear dichroism study of field-manipulated canted antiferromagnetism in epitaxial $\alpha$-{Fe}$_{2}${O}$_{3}$ films},
	volume = {5},
	issn = {2475-9953},
	url = {https://link.aps.org/doi/10.1103/PhysRevMaterials.5.044408},
	doi = {10.1103/PhysRevMaterials.5.044408},
	number = {4},
	urldate = {2025-09-25},
	journal = {Physical Review Materials},
	author = {Suturin, Sergey M. and Korovin, Alexander M. and Gastev, Sergey V. and Dvortsova, Polina A. and Volkov, Mikhail P. and Valvidares, Manuel and Sokolov, Nikolai S.},
	month = apr,
	year = {2021},
	pages = {044408},
}

@article{Takegami25,
  title = {{Circular Dichroism in Resonant Inelastic X-Ray Scattering: Probing Altermagnetic Domains in MnTe}},
  author = {Takegami, D. and Aoyama, T. and Okauchi, T. and Yamaguchi, T. and Tippireddy, S. and Agrestini, S. and Garc\'{\i}a-Fern\'andez, M. and Mizokawa, T. and Ohgushi, K. and Zhou, Ke-Jin and Chaloupka, J. and Kune\ifmmode \check{s}\else \v{s}\fi{}, J. and Hariki, A. and Suzuki, H.},
  journal = {Phys. Rev. Lett.},
  volume = {135},
  issue = {19},
  pages = {196502},
  numpages = {6},
  year = {2025},
  month = {Nov},
  publisher = {American Physical Society},
  doi = {10.1103/512v-n5f9},
  url = {https://link.aps.org/doi/10.1103/512v-n5f9}
}

@article{Luo26,
	title = {Symmetry-driven giant magneto-optical {K}err effects in altermagnetic insulator},
	url = {http://iopscience.iop.org/article/10.1088/0256-307X/43/2/020713},
	journal = {Chinese Physics Letters},
	author = {Luo, Jiaxin and Zhou, Xiaodong and Liang, Jinxuan and Wang, Ledong and Zhou, Qiuyun and Jiang, Yong and Wang, Wenhong and Yao, Yugui and Yang, Luyi and Jiang, Wanjun},
	year = {2026},
}

@Article{Mayr21,
AUTHOR = {Mayr, Sina and Finizio, Simone and Reuteler, Joakim and Stutz, Stefan and Dubs, Carsten and Weigand, Markus and Hrabec, Aleš and Raabe, Jörg and Wintz, Sebastian},
TITLE = {{Xenon Plasma Focused Ion Beam Milling for Obtaining Soft X-ray Transparent Samples}},
JOURNAL = {Crystals},
VOLUME = {11},
YEAR = {2021},
NUMBER = {5},
ARTICLE-NUMBER = {546},
URL = {https://www.mdpi.com/2073-4352/11/5/546},
ISSN = {2073-4352},
DOI = {10.3390/cryst11050546}
}

@Article{Weigand22,
AUTHOR = {Weigand, Markus and Wintz, Sebastian and Gräfe, Joachim and Noske, Matthias and Stoll, Hermann and Van Waeyenberge, Bartel and Schütz, Gisela},
TITLE = {{Time{M}axyne: A Shot-Noise Limited, Time-{R}esolved Pump-and-{P}robe Acquisition System Capable of 50 {GH}z Frequencies for Synchrotron-{B}ased X-ray Microscopy}},
JOURNAL = {Crystals},
VOLUME = {12},
YEAR = {2022},
NUMBER = {8},
ARTICLE-NUMBER = {1029},
URL = {https://www.mdpi.com/2073-4352/12/8/1029},
ISSN = {2073-4352},
DOI = {10.3390/cryst12081029}
}

@article{Moriya60,
  title = {Anisotropic Superexchange Interaction and Weak Ferromagnetism},
  author = {Moriya, T\^oru},
  journal = {Phys. Rev.},
  volume = {120},
  issue = {1},
  pages = {91--98},
  numpages = {0},
  year = {1960},
  month = {Oct},
  publisher = {American Physical Society},
  doi = {10.1103/PhysRev.120.91},
  url = {https://link.aps.org/doi/10.1103/PhysRev.120.91}
}

@article{Dzyaloshinsky58,
	title = {A thermodynamic theory of “weak” ferromagnetism of antiferromagnetics},
	volume = {4},
	issn = {0022-3697},
	url = {https://www.sciencedirect.com/science/article/pii/0022369758900763},
	doi = {https://doi.org/10.1016/0022-3697(58)90076-3},
	number = {4},
	journal = {Journal of Physics and Chemistry of Solids},
	author = {Dzyaloshinsky, I.},
	year = {1958},
	pages = {241--255},
}

@article{Jani21,
	title = {Antiferromagnetic half-skyrmions and bimerons at room temperature},
	volume = {590},
	issn = {1476-4687},
	url = {https://doi.org/10.1038/s41586-021-03219-6},
	doi = {10.1038/s41586-021-03219-6},
	number = {7844},
	journal = {Nature},
	author = {Jani, Hariom and Lin, Jheng-Cyuan and Chen, Jiahao and Harrison, Jack and Maccherozzi, Francesco and Schad, Jonathon and Prakash, Saurav and Eom, Chang-Beom and Ariando, A. and Venkatesan, T. and Radaelli, Paolo G.},
	month = feb,
	year = {2021},
	pages = {74--79},
}

@article{Morin50,
  title = {Magnetic Susceptibility of $\alpha$-{Fe}$_{2}${O}$_{3}$ and $\alpha$-{Fe}$_{2}${O}$_{3}$ with Added Titanium},
  author = {Morin, F. J.},
  journal = {Phys. Rev.},
  volume = {78},
  issue = {6},
  pages = {819--820},
  numpages = {0},
  year = {1950},
  month = {Jun},
  publisher = {American Physical Society},
  doi = {10.1103/PhysRev.78.819.2},
  url = {https://link.aps.org/doi/10.1103/PhysRev.78.819.2}
}

@article{Feng22,
	title = {An anomalous {Hall} effect in altermagnetic ruthenium dioxide},
	volume = {5},
	issn = {2520-1131},
	url = {https://doi.org/10.1038/s41928-022-00866-z},
	doi = {10.1038/s41928-022-00866-z},
	number = {11},
	journal = {Nature Electronics},
	author = {Feng, Zexin and Zhou, Xiaorong and Šmejkal, Libor and Wu, Lei and Zhu, Zengwei and Guo, Huixin and González-Hernández, Rafael and Wang, Xiaoning and Yan, Han and Qin, Peixin and Zhang, Xin and Wu, Haojiang and Chen, Hongyu and Meng, Ziang and Liu, Li and Xia, Zhengcai and Sinova, Jairo and Jungwirth, Tomáš and Liu, Zhiqi},
	month = nov,
	year = {2022},
	pages = {735--743},
}

@misc{Mazin23,
  title   = {{Induced Monolayer Altermagnetism in {MnP(S,Se)}$_3$ and {FeSe}}},
  author  = {Mazin, Igor and Gonz{\'a}lez-Hern{\'a}ndez, Rafael and {\v{S}}mejkal, Libor},
  year    = {2023},
  eprint  = {2309.02355},
  archivePrefix = {arXiv},
  primaryClass  = {cond-mat.mes-hall},
  doi     = {10.48550/arXiv.2309.02355}
}

@article{Tschirner23,
	title = {Saturation of the anomalous {Hall} effect at high magnetic fields in altermagnetic {RuO}$_{2}$},
	volume = {11},
	issn = {2166-532X},
	url = {https://doi.org/10.1063/5.0160335},
	doi = {10.1063/5.0160335},
	number = {10},
	journal = {APL Materials},
	author = {Tschirner, Teresa and Keßler, Philipp and Gonzalez Betancourt, Ruben Dario and Kotte, Tommy and Kriegner, Dominik and Büchner, Bernd and Dufouleur, Joseph and Kamp, Martin and Jovic, Vedran and Smejkal, Libor and Sinova, Jairo and Claessen, Ralph and Jungwirth, Tomas and Moser, Simon and Reichlova, Helena and Veyrat, Louis},
	month = oct,
	year = {2023},
	pages = {101103},
}

@article{Wang23,
  author  = {Wang, Meng and Tanaka, Katsuhiro and Sakai, Shiro and Wang, Ziqian and Deng, Ke and Lyu, Yingjie and Li, Cong and Tian, Di and Shen, Shengchun and Ogawa, Naoki and Kanazawa, Naoya and Yu, Pu and Arita, Ryotaro and Kagawa, Fumitaka},
  title   = {Emergent zero-field anomalous Hall effect in a reconstructed rutile antiferromagnetic metal},
  journal = {Nature Communications},
  volume  = {14},
  number  = {1},
  pages   = {8240},
  year    = {2023},
  doi     = {10.1038/s41467-023-43962-0}
}

@article{Jeong25,
author = {Seung Gyo Jeong  and Seungjun Lee  and Bonnie Lin  and Zhifei Yang  and In Hyeok Choi  and Jin Young Oh  and Sehwan Song  and Seung wook Lee  and Sreejith Nair  and Rashmi Choudhary  and Juhi Parikh  and Sungkyun Park  and Woo Seok Choi  and Jong Seok Lee  and James M. LeBeau  and Tony Low  and Bharat Jalan },
title = {Metallicity and anomalous Hall effect in epitaxially strained, atomically thin {RuO}$_{2}$ films},
journal = {Proceedings of the National Academy of Sciences},
volume = {122},
number = {24},
pages = {e2500831122},
year = {2025},
doi = {10.1073/pnas.2500831122},
URL = {https://www.pnas.org/doi/abs/10.1073/pnas.2500831122}
}

@article{Chu25,
  title = {Third-Order Nonlinear Hall Effect in Altermagnet {RuO}$_{2}$},
  author = {Chu, R. Y. and Han, L. and Gong, Z. H. and Fu, X. Z. and Bai, H. and Liang, S. X. and Chen, C. and Cheong, S-W. and Zhang, Y. Y. and Liu, J. W. and Wang, Y. Y. and Pan, F. and Lu, H. Z. and Song, C.},
  journal = {Phys. Rev. Lett.},
  volume = {135},
  issue = {21},
  pages = {216703},
  numpages = {8},
  year = {2025},
  month = {Nov},
  publisher = {American Physical Society},
  doi = {10.1103/rv1n-vr4p},
  url = {https://link.aps.org/doi/10.1103/rv1n-vr4p}
}

@article{Kluczyk24,
  title = {Coexistence of anomalous Hall effect and weak magnetization in a nominally collinear antiferromagnet {MnTe}},
  author = {Kluczyk, K. P. and Gas, K. and Grzybowski, M. J. and Skupi\ifmmode \acute{n}\else \'{n}\fi{}ski, P. and Borysiewicz, M. A. and F\k{a}s, T. and Suffczy\ifmmode \acute{n}\else \'{n}\fi{}ski, J. and Domagala, J. Z. and Grasza, K. and Mycielski, A. and Baj, M. and Ahn, K. H. and V\'yborn\'y, K. and Sawicki, M. and Gryglas-Borysiewicz, M.},
  journal = {Phys. Rev. B},
  volume = {110},
  issue = {15},
  pages = {155201},
  numpages = {12},
  year = {2024},
  month = {Oct},
  publisher = {American Physical Society},
  doi = {10.1103/PhysRevB.110.155201},
  url = {https://link.aps.org/doi/10.1103/PhysRevB.110.155201}
}

@article{Reichlova24,
  author  = {Reichlova, Helena and Lopes Seeger, Rafael and González-Hernández, Rafael and Kounta, Ismaila and Schlitz, Richard and Kriegner, Dominik and Ritzinger, Philipp and Lammel, Michaela and Leiviskä, Miina and Birk Hellenes, Anna and Olejník, Kamil and Petřiček, Vaclav and Doležal, Petr and Horak, Lukas and Schmoranzerova, Eva and Badura, Antonín and Bertaina, Sylvain and Thomas, Andy and Baltz, Vincent and Michez, Lisa and Sinova, Jairo and Goennenwein, Sebastian T. B. and Jungwirth, Tomáš and Šmejkal, Libor},
  title   = {Observation of a spontaneous anomalous Hall response in the {Mn\textsubscript{5}Si\textsubscript{3}} d-wave altermagnet candidate},
  journal = {Nature Communications},
  volume  = {15},
  number  = {1},
  pages   = {4961},
  year    = {2024},
  doi     = {10.1038/s41467-024-48493-w}
}

@article{GonzalezBetancourt24,
  author  = {Gonzalez Betancourt, Ruben Dario and Zubáč, Jan and Geishendorf, Kevin and Ritzinger, Philipp and Růžičková, Barbora and Kotte, Tommy and Železný, Jakub and Olejník, Kamil and Springholz, Gunther and Büchner, Bernd and Thomas, Andy and Výborný, Karel and Jungwirth, Tomas and Reichlová, Helena and Kriegner, Dominik},
  title   = {Anisotropic magnetoresistance in altermagnetic {MnTe}},
  journal = {npj Spintronics},
  volume  = {2},
  number  = {1},
  pages   = {45},
  year    = {2024},
  doi     = {10.1038/s44306-024-00046-z}
}

@article{
Han24,
author = {Lei Han  and Xizhi Fu  and Rui Peng  and Xingkai Cheng  and Jiankun Dai  and Liangyang Liu  and Yidian Li  and Yichi Zhang  and Wenxuan Zhu  and Hua Bai  and Yongjian Zhou  and Shixuan Liang  and Chong Chen  and Qian Wang  and Xianzhe Chen  and Luyi Yang  and Yang Zhang  and Cheng Song  and Junwei Liu  and Feng Pan },
title = {Electrical 180° switching of {N}éel vector in spin-splitting antiferromagnet},
journal = {Science Advances},
volume = {10},
number = {4},
pages = {eadn0479},
year = {2024},
doi = {10.1126/sciadv.adn0479},
URL = {https://www.science.org/doi/abs/10.1126/sciadv.adn0479}}

@article{Iguchi25,
  title = {Magneto-optical spectra of an organic antiferromagnet as a candidate for an altermagnet},
  author = {Iguchi, Satoshi and Kobayashi, Hiroki and Ikemoto, Yuka and Furukawa, Tetsuya and Itoh, Hirotake and Iwai, Shinichiro and Moriwaki, Taro and Sasaki, Takahiko},
  journal = {Phys. Rev. Res.},
  volume = {7},
  issue = {3},
  pages = {033026},
  numpages = {13},
  year = {2025},
  month = {Jul},
  publisher = {American Physical Society},
  doi = {10.1103/nnz3-tq7y},
  url = {https://link.aps.org/doi/10.1103/nnz3-tq7y}
}

@article{Pan26,
  title = {Experimental Evidence of N\'eel-Order-Driven Magneto-optical Kerr Effect in an Altermagnetic Insulator},
  author = {Pan, Haolin and Xiao, Rui-Chun and Han, Jiahao and Zhu, Hongxing and Li, Junxue and Niu, Qian and Gao, Yang and Hou, Dazhi},
  journal = {Phys. Rev. Lett.},
  volume = {136},
  issue = {3},
  pages = {036701},
  numpages = {7},
  year = {2026},
  month = {Jan},
  publisher = {American Physical Society},
  doi = {10.1103/q8ym-l2zt},
  url = {https://link.aps.org/doi/10.1103/q8ym-l2zt}
}

@article{Biniskos25,
  author  = {Biniskos, Nikolaos and dos Santos Dias, Manuel and Agrestini, Stefano and Sviták, David and Zhou, Ke-Jin and Pospíšil, Jiří and Čermák, Petr},
  title   = {Systematic mapping of altermagnetic magnons by resonant inelastic X-ray circular dichroism},
  journal = {Nature Communications},
  volume  = {16},
  number  = {1},
  pages   = {9311},
  year    = {2025},
  doi     = {10.1038/s41467-025-64322-0}
}

@article{Liu24,
  title = {Chiral Split Magnon in Altermagnetic {MnTe}},
  author = {Liu, Zheyuan and Ozeki, Makoto and Asai, Shinichiro and Itoh, Shinichi and Masuda, Takatsugu},
  journal = {Phys. Rev. Lett.},
  volume = {133},
  issue = {15},
  pages = {156702},
  numpages = {6},
  year = {2024},
  month = {Oct},
  publisher = {American Physical Society},
  doi = {10.1103/PhysRevLett.133.156702},
  url = {https://link.aps.org/doi/10.1103/PhysRevLett.133.156702}
}

@article{Yang25,
  author  = {Yang, Guowei and Li, Zhanghuan and Yang, Sai and Li, Jiyuan and Zheng, Hao and Zhu, Weifan and Pan, Ze and Xu, Yifu and Cao, Saizheng and Zhao, Wenxuan and Jana, Anupam and Zhang, Jiawen and Ye, Mao and Song, Yu and Hu, Lun-Hui and Yang, Lexian and Fujii, Jun and Vobornik, Ivana and Shi, Ming and Yuan, Huiqiu and Zhang, Yongjun and Xu, Yuanfeng and Liu, Yang},
  title   = {Three-dimensional mapping of the altermagnetic spin splitting in {CrSb}},
  journal = {Nature Communications},
  volume  = {16},
  number  = {1},
  pages   = {1442},
  year    = {2025},
  doi     = {10.1038/s41467-025-56647-7}
}

@article{Zeng24,
author = {Zeng, Meng and Zhu, Ming-Yuan and Zhu, Yu-Peng and Liu, Xiang-Rui and Ma, Xiao-Ming and Hao, Yu-Jie and Liu, Pengfei and Qu, Gexing and Yang, Yichen and Jiang, Zhicheng and Yamagami, Kohei and Arita, Masashi and Zhang, Xiaoqian and Shao, Tian-Hao and Dai, Yue and Shimada, Kenya and Liu, Zhengtai and Ye, Mao and Huang, Yaobo and Liu, Qihang and Liu, Chang},
title = {Observation of Spin Splitting in Room-Temperature Metallic Antiferromagnet {CrSb}},
journal = {Advanced Science},
volume = {11},
number = {43},
pages = {2406529},
doi = {https://doi.org/10.1002/advs.202406529},
url = {https://advanced.onlinelibrary.wiley.com/doi/abs/10.1002/advs.202406529},
year = {2024}
}

@article{Li25,
  author  = {Li, Cong and Hu, Mengli and Li, Zhilin and Wang, Yang and Chen, Wanyu and Thiagarajan, Balasubramanian and Leandersson, Mats and Polley, Craig and Kim, Timur and Liu, Hui and Fulga, Cosma and Vergniory, Maia G. and Janson, Oleg and Tjernberg, Oscar and van den Brink, Jeroen},
  title   = {Topological {Weyl} altermagnetism in {CrSb}},
  journal = {Communications Physics},
  volume  = {8},
  number  = {1},
  pages   = {311},
  year    = {2025},
  doi     = {10.1038/s42005-025-02232-9}
}

@article{Lu25,
  author  = {Lu, Wenlong and Feng, Shiyu and Wang, Yuzhi and Chen, Dong and Lin, Zihan and Liang, Xin and Liu, Siyuan and Feng, Wanxiang and Yamagami, Kohei and Liu, Junwei and Felser, Claudia and Wu, Quansheng and Ma, Junzhang},
  title   = {Signature of Topological Surface Bands in Altermagnetic {Weyl} Semimetal {CrSb}},
  journal = {Nano Letters},
  volume  = {25},
  number  = {18},
  pages   = {7343--7350},
  year    = {2025},
  doi     = {10.1021/acs.nanolett.5c00482},
  publisher = {American Chemical Society}
}

@article{Santhosh25,
author = {Santhosh, Sandra and Corbae, Paul and Yánez-Parreño, Wilson J. and Ghosh, Supriya and Jensen, Christopher J. and Fedorov, Alexei V. and Hashimoto, Makoto and Lu, Donghui and Borchers, Julie A. and Grutter, Alexander J. and Charlton, Timothy R. and Islam, Saurav and Golovanova, Diana and Zhao, Yufei and Tauraso, Aria and Richardella, Anthony and Yan, Binghai and Mkhoyan, K. Andre and Palmstrøm, Christopher J. and Ou, Yongxi and Samarth, Nitin},
title = {Altermagnetic Band Splitting in 10 nm Epitaxial {CrSb} Thin Films},
journal = {Advanced Materials},
volume = {37},
number = {47},
pages = {e08977},
doi = {https://doi.org/10.1002/adma.202508977},
url = {https://advanced.onlinelibrary.wiley.com/doi/abs/10.1002/adma.202508977},
year = {2025}
}

@article{Kuiper93,
  title = {X-ray magnetic dichroism of antiferromagnet {Fe}$_{2}${O}$_{3}$: The orientation of magnetic moments observed by Fe 2p x-ray absorption spectroscopy},
  author = {Kuiper, Pieter and Searle, Barry G. and Rudolf, Petra and Tjeng, L. H. and Chen, C. T.},
  journal = {Phys. Rev. Lett.},
  volume = {70},
  issue = {10},
  pages = {1549--1552},
  numpages = {0},
  year = {1993},
  month = {Mar},
  publisher = {American Physical Society},
  doi = {10.1103/PhysRevLett.70.1549},
  url = {https://link.aps.org/doi/10.1103/PhysRevLett.70.1549}
}

@article{Smejkal23,
  title = {{Chiral Magnons in Altermagnetic RuO$_{2}$}},
  author = {\ifmmode \check{S}\else \v{S}\fi{}mejkal, Libor and Marmodoro, Alberto and Ahn, Kyo-Hoon and Gonz\'alez-Hern\'andez, Rafael and Turek, Ilja and Mankovsky, Sergiy and Ebert, Hubert and D'Souza, Sunil W. and \ifmmode \check{S}\else \v{S}\fi{}ipr, Ond\ifmmode \check{r}\else \v{r}\fi{}ej and Sinova, Jairo and Jungwirth, Tom{\'a}\v{s}},
  journal = {Phys. Rev. Lett.},
  volume = {131},
  issue = {25},
  pages = {256703},
  numpages = {6},
  year = {2023},
  month = {Dec},
  publisher = {American Physical Society},
  doi = {10.1103/PhysRevLett.131.256703},
  url = {https://link.aps.org/doi/10.1103/PhysRevLett.131.256703}
}

@article{Marzin21,
author = {Igor I. Mazin  and Klaus Koepernik  and Michelle D. Johannes  and Rafael González-Hernández  and Libor Šmejkal },
title = {Prediction of unconventional magnetism in doped {FeSb$_{2}$}},
journal = {Proceedings of the National Academy of Sciences},
volume = {118},
number = {42},
pages = {e2108924118},
year = {2021},
doi = {10.1073/pnas.2108924118},
URL = {https://www.pnas.org/doi/abs/10.1073/pnas.2108924118},
eprint = {https://www.pnas.org/doi/pdf/10.1073/pnas.2108924118}}

@article{Zhou25,
  author  = {Zhou, Zhiyuan and Cheng, Xingkai and Hu, Mengli and Chu, Ruiyue and Bai, Hua and Han, Lei and Liu, Junwei and Pan, Feng and Song, Cheng},
  title   = {Manipulation of the altermagnetic order in {CrSb} via crystal symmetry},
  journal = {Nature},
  year    = {2025},
  volume  = {638},
  number  = {8051},
  pages   = {645--650},
  doi     = {10.1038/s41586-024-08436-3},
  url     = {https://doi.org/10.1038/s41586-024-08436-3},
  issn    = {1476-4687}
}

@article{Takagi25,
  author  = {Takagi, Rina and Hirakida, Ryosuke and Settai, Yuki and Oiwa, Rikuto and Takagi, Hirotaka and Kitaori, Aki and Yamauchi, Kensei and Inoue, Hiroki and Yamaura, Jun-ichi and Nishio-Hamane, Daisuke and Itoh, Shinichi and Aji, Seno and Saito, Hiraku and Nakajima, Taro and Nomoto, Takuya and Arita, Ryotaro and Seki, Shinichiro},
  title   = {Spontaneous {H}all effect induced by collinear antiferromagnetic order at room temperature},
  journal = {Nature Materials},
  year    = {2025},
  volume  = {24},
  number  = {1},
  pages   = {63--68},
  doi     = {10.1038/s41563-024-02058-w},
  url     = {https://doi.org/10.1038/s41563-024-02058-w},
  issn    = {1476-4660}
}

@misc{Xie25,
  title        = {X‐ray magnetic circular dichroism of altermagnet $\alpha$‐{Fe}$_2${O}$_3$ based on multiplet ligand‐field theory using {W}annier orbitals},
  author       = {Ruiwen Xie and Hamza Zerdoumi and Hongbin Zhang},
  year         = {2025},
  eprint       = {2512.11664},
  archivePrefix= {arXiv},
  primaryClass = {cond-mat.mtrl-sci},
  doi          = {10.48550/arXiv.2512.11664},
  note         = {arXiv preprint arXiv:2512.11664}
}

@misc{Ishii26,
  title        = {Altermagnetic {XMCD} in {Hematite} {Distinct} from {Weak} {Ferromagnetic} {Contributions}},
  author       = {Ishii, Y. and Sasabe, N. and Yamasaki, Y.},
  year         = {2026},
  eprint       = {2603.00442},
  archivePrefix= {arXiv},
  primaryClass = {cond-mat.str-el},
  doi          = {10.48550/arXiv.2603.00442}
}

@article{Tan24,
  author  = {Tan, Anthony K. C. and Jani, Hariom and Högen, Maximilian and Mechnich, Annika and Feuer, Matthew S. G. and Knowles, Helena S. and Ariando and Radaelli, Paolo G. and Atatüre, Mete},
  title   = {Revealing emergent magnetic charge in an antiferromagnet with diamond quantum magnetometry},
  journal = {Nature Materials},
  volume  = {23},
  pages   = {205--211},
  year    = {2024},
  doi     = {10.1038/s41563-023-01737-4}
}

\clearpage
\renewcommand{\figurename}{Extended Data Fig.}
\setcounter{figure}{0}
\begin{figure*}[h]
\begin{center}
\includegraphics[width=1.0\columnwidth]{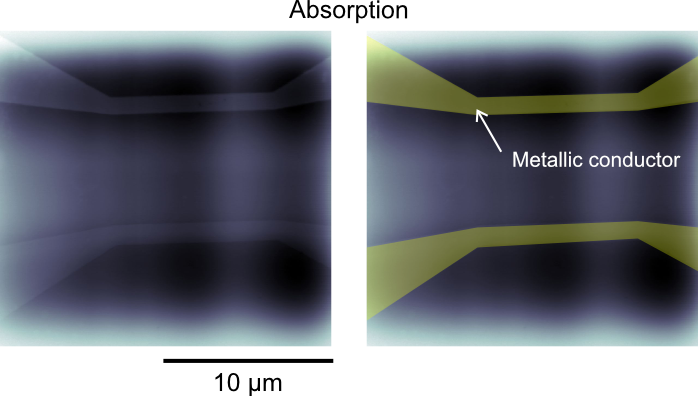}
\end{center}
\caption{Overview absorption images from the STXM experiments. The metallic conductor regions are highlighted in yellow as a guide to the eye.}
\label{ex_fig_1}
\end{figure*}

\begin{figure*}[h]
\begin{center}
\includegraphics[width=0.5\columnwidth]{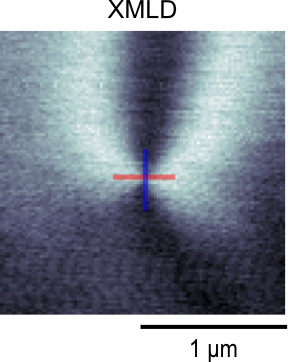}
\end{center}
\caption{XMLD image at 724.5~eV. The lines used for the line profiles in Fig.~\ref{fig_4}\textbf{d} and \textbf{e} are indicated.}
\label{ex_fig_2}
\end{figure*}

\end{document}


\title{Supplementary Information: \\
Nanoscale imaging of spin textures \\with locally varying altermagnetic response in $\alpha$-Fe$_2$O$_3$}

\author{R. Yamamoto}
\thanks{These authors contributed equally to this work.}
\affiliation{Max Planck Institute for Chemical Physics of Solids, 01187 Dresden, Germany}
\affiliation{International Institute for Sustainability with Knotted Chiral Meta Matter (WPI-SKCM$^{2}$), Hiroshima University, Hiroshima 739-8526, Japan}

\author{S. Mayr}
\thanks{These authors contributed equally to this work.}
\affiliation{Paul Scherrer Institut, 5232 Villigen PSI, Switzerland}
\affiliation{Laboratory for Mesoscopic Systems, Department of Materials, ETH Z\"urich
, 8093 Z\"urich, Switzerland}

\author{A. Hariki}
\affiliation{Department of Physics and Electronics, Graduate School of Engineering,
Osaka Metropolitan University, 1-1 Gakuen-cho, Nakaku, Sakai, Osaka 599-8531, Japan}

\author{S. Finizio}
\affiliation{Paul Scherrer Institut, 5232 Villigen PSI, Switzerland}

\author{K. Sakurai}
\affiliation{Department of Physics and Electronics, Graduate School of Engineering,
Osaka Metropolitan University, 1-1 Gakuen-cho, Nakaku, Sakai, Osaka 599-8531, Japan}

\author{E. Weschke}
\affiliation{Helmholtz-Zentrum Berlin f\"ur Materialien und Energie GmbH, 12409 Berlin, Germany}

\author{K. Litzius}
\affiliation{Universität of Augsburg, 86150 Augsburg, Germany}

\author{M. T. Birch}
\affiliation{RIKEN Center for Emergent Matter Science (CEMS), Wako, Saitama 351-0198, Japan}

\author{L. A. Turnbull}
\affiliation{Max Planck Institute for Chemical Physics of Solids, 01187 Dresden, Germany}
\affiliation{Diamond Light Source, Didcot, Oxfordshire, OX11 0DE, United Kingdom}

\author{E. Zhakina}
\affiliation{Max Planck Institute for Chemical Physics of Solids, 01187 Dresden, Germany}

\author{M. Di Pietro Mart\'inez}
\affiliation{Max Planck Institute for Chemical Physics of Solids, 01187 Dresden, Germany}
\affiliation{International Institute for Sustainability with Knotted Chiral Meta Matter (WPI-SKCM$^{2}$), Hiroshima University, Hiroshima 739-8526, Japan}

\author{J. Reuteler}
\affiliation{ETH Z\"urich, 8093 Z\"urich, Switzerland}

\author{F. Schulz}
\affiliation{Max-PLanck-Institut f\"ur Intelligente Systeme, 70569 Sttutgart, Germany}

\author{M. Weigand}
\affiliation{Helmholtz-Zentrum Berlin f\"ur Materialien und Energie GmbH, 12409 Berlin, Germany}

\author{J. Raabe}
\affiliation{Paul Scherrer Institut, 5232 Villigen PSI, Switzerland}

\author{G. Sch\"utz}
\affiliation{Max-PLanck-Institut f\"ur Intelligente Systeme, 70569 Sttutgart, Germany}

\author{S. S. P. K. Arekapudi}
\affiliation{Helmholtz-Zentrum Dresden-Rossendorf, 01328 Dresden, Germany}
\affiliation{Technische Universität Chemnitz, 09111 Chemnitz, Germany}

\author{O. Hellwig}
\affiliation{Helmholtz-Zentrum Dresden-Rossendorf, 01328 Dresden, Germany}
\affiliation{Technische Universität Chemnitz, 09111 Chemnitz, Germany}

\author{W. H. Campos}
\affiliation{Max Planck Institute for the Physics of Complex Systems, 01187 Dresden, Germany}

\author{L. \v{S}mejkal}
\affiliation{Max Planck Institute for the Physics of Complex Systems, 01187 Dresden, Germany}

\author{J. Kune\v{s}}
\affiliation{Department of Condensed Matter Physics, Faculty of Science, Masaryk University, Kotl\'{a}\v{r}sk\'{a}  2, 61137 Brno, Czech Republic}

\author{C. Donnelly}
\affiliation{Max Planck Institute for Chemical Physics of Solids, 01187 Dresden, Germany}
\affiliation{International Institute for Sustainability with Knotted Chiral Meta Matter (WPI-SKCM$^{2}$), Hiroshima University, Hiroshima 739-8526, Japan}

\author{S.~Wintz}
\affiliation{Helmholtz-Zentrum Berlin f\"ur Materialien und Energie GmbH, 12409 Berlin, Germany}

\date{\today}

\maketitle

\subsubsection{Characterization of single crystals}
The single crystallinity and lateral crystallographic orientation of Sample \#1 was confirmed by Laue measurements.
The Laue pattern of Sample \#1 is displayed in Supplementary Figure \ref{si_fig_1}.

\begin{figure}[h]
\begin{center}
\includegraphics[width=0.5\columnwidth]{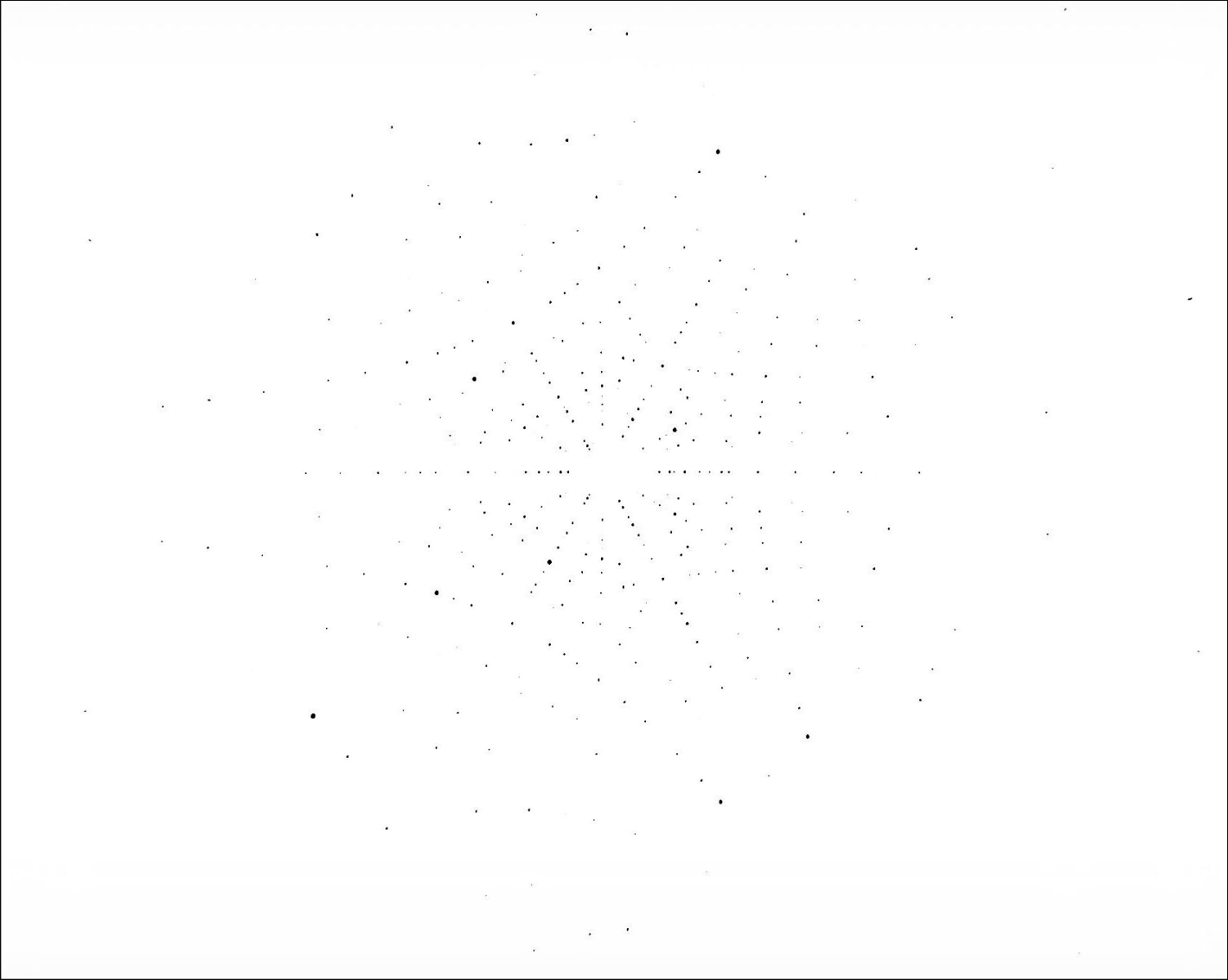}
\end{center}
\caption{Laue diffraction pattern of single crystalline $\alpha$-Fe$_{2}$O$_{3}$ (Sample \#1).
} 
\label{si_fig_1}
\end{figure}

\section{Symmetry analysis}
According to our spin group theory analysis, $\alpha$-Fe$_2$O$_3$ below the Néel temperature $T_N \approx 955~$K is described by the spin point group $\bar{3}m^{\infty m}$ and thus belongs to the $g$-wave symmetry altermagnets~\cite{Smejkal22a}. As illustrated in Supplementary Figure \ref{si_fig_4}, the four spin sublattices in a rhombohedral unit cell exhibit, in addition to antiparallel dipolar order, ferroically ordered $g$-wave spin densities~\cite{Bhowal24,Verbeek24,Jaeschke25}. This ferroic ordering gives rise to a $g$-wave spin-polarized electronic structure, shown in Fig.~1\textbf{b}. Three vertical spin-degenerate nodal surfaces are symmetry-enforced to be flat. In contrast, the fourth in-plane nodal plane is not symmetry constrained and, owing to its curved shape, forms additional accidental crossings in the $k_z = 0$ plane~\cite{Hoyer25}. This nodal topology \cite{Mazin23} is therefore distinct from that of the $g$-wave altermagnet $\alpha$-MnTe, which exhibits a flat horizontal nodal surface.


\begin{figure}[h]
\begin{center}
\includegraphics[width=0.45\columnwidth]{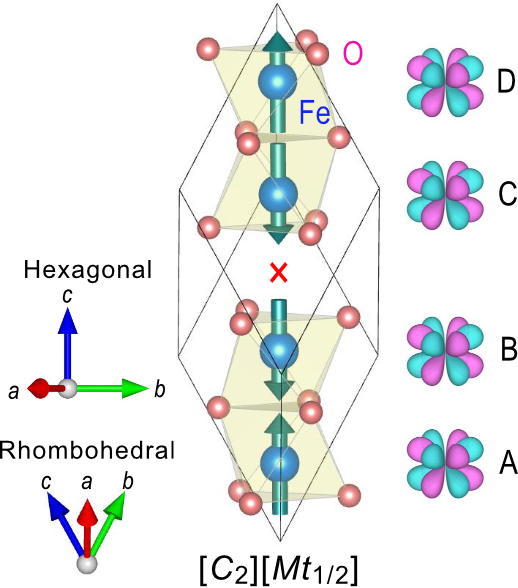}
\end{center}
\caption{Magnetic structure of $\alpha$-Fe$_2$O$_3$ in a rhombohedral unit cell. 
Antiferromagnetic ordering on the four sublattices (A-D) breaks  space-time inversion ($\mathcal{P}\mathcal{T}$) symmetry with respect to the inversion center highlighted by the red cross. The collinear spin order, together with the oxygen octahedra, stabilizes a $g$-wave altermagnetic order that exhibits ferroic alignment across the four spin sublattices, as illustrated in the left column.}
\label{si_fig_4}
\end{figure}

\section{Magnetic structures of $\alpha$-Fe$_{2}$O$_{3}$ and expected XMCD/XMLD contrast}
\begin{figure*}[]
\begin{center}
\includegraphics{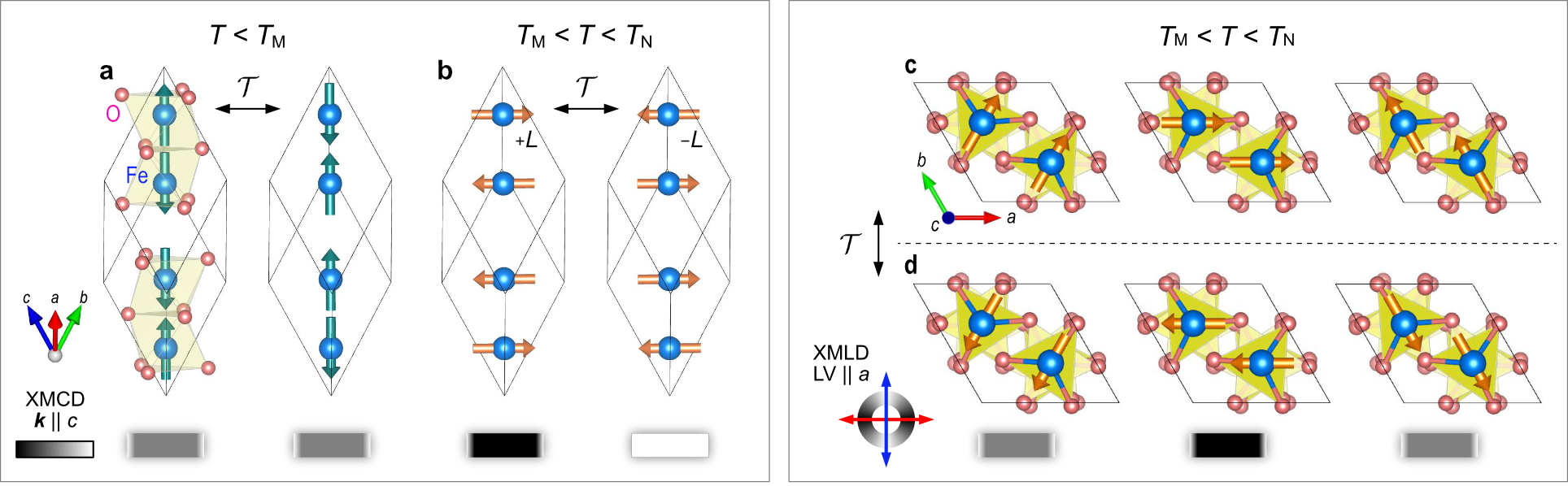}
\end{center}
\caption{Crystal and magnetic structures of $\alpha$-Fe$_{2}$O$_{3}$ and the expected XMCD and XMLD contrast for the experimental geometry used in this work.
\textbf{a}. Magnetic structure below the Morin transition temperature ($T<T_\mathrm{M}$), where magnetic moments are aligned along the 
$c$-axis. XMCD measured with the X-ray wave vector $\bm{k}\,||\,c$ is expected to vanish.
\textbf{b}. Spin configuration in the range $T_\mathrm{M}<T<T_\mathrm{N}$, where spins lie in the basal plane, resulting in a finite XMCD contrast, depending on $\mathbf{L}$.
\textbf{c,d} Expected XMLD contrast for different in-plane orientations of the N\'eel vector.
} 
\label{si_fig_2}
\end{figure*}

Supplementary Figure \ref{si_fig_2} schematically summarizes the magnetic structure of $\alpha$-Fe$_{2}$O$_{3}$ and the XMCD and XMLD contrast expected for the experimental geometry used in this work.
For temperatures below the Morin transition ($T<T_\mathrm{M}$), the spins are collinearly aligned along the $c$-axis. In this phase, XMCD measured with the X-ray wave vector $\bm{k}\,||\,c$-axis vanishes.
In the temperature range $T_\mathrm{M}<T<T_\mathrm{N}$, the spins rotate into the basal plane, leading to a finite XMCD signal, depending on the N\'eel vector azimuthal orientation and sign ($\pm L$).

In contrast, XMLD is sensitive to the orientation of the N\'eel vector with respect to the linear polarization direction of the incident X-rays, enabling the identification of antiferromagnetic domains.  

\section{Analysis of scanning transmission X-ray microscopy data}


\subsection{Calibrating thickness variation of absorption image}
To improve visibility of the magnetic contrast, the contribution from the metallic lead stripline in the absorption image shown in Supplementary Fig.~\ref{si_fig_3}\textbf{a} was calibrated using the following procedure and obtained the image displayed in Fig.~3\textbf{e}.
In this procedure, an absorption image acquired at low temperature, where no magnetic features are present, was used as reference.
The reference image was rescaled and spatially aligned to the corresponding target image, and the intensities were normalized using the signal at the central region of the field of view, as shown in Supplementary Fig.~\ref{si_fig_3}\textbf{b}.
The normalized target absorption image $I$ was then calibrated by the reference image $I_\textrm{reference}$ according to
\begin{equation}
I'
= -\frac{I}{I_\textrm{reference}},
\label{eq:si_eq_2}
\end{equation}

where the minus sign is introduced to ensure consistency with the contrast convention used for the calculations displayed in Fig.~3\textbf{c}.
The absorption image obtained using this procedure is shown in Fig.~3\textbf{e}.

\begin{figure}[h]
\begin{center}
\includegraphics[width=0.75\columnwidth]{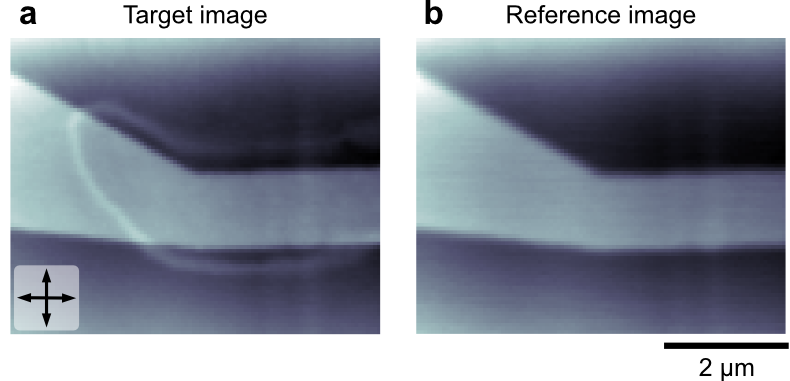}
\end{center}
\caption{\textbf{a} Absorption image prior to background removal. The white region is the metallic lead stripline.
\textbf{b} Reference absorption image acquired at low temperature, which is rescaled, spatially aligned, and intensity-normalized using the signal at the central region of the field of view.}
\label{si_fig_3}
\end{figure}


\subsection{Quantitative analysis of dichroic signal}
To quantitatively evaluate the XMCD obtained from the transmission experiment, we use the following definition,
\begin{equation}
\mathrm{XMCD} = 100 \times \frac{-\ln I^{-} + \ln I^{+}_{\mathrm{norm}}}{A_{\mathrm{max}}}.
\label{eq:si_eq_3}
\end{equation}
Here, $A_{\mathrm{max}}$ denotes the absorption peak value at the $L_3$ edge. Transmission images for left- and right-circularly polarized X-rays, $I^{\pm}$, were acquired as a function of photon energy, and $I^{+}_{\mathrm{norm}}$ was obtained by the normalization procedure described in the Methods section.

In the transmission experiment performed at the MAXYMUS end station, the large sample thickness prevents a reliable determination of the $L_3$-edge absorption peak from the transmission data. Instead, $A_{\mathrm{max}}$ is estimated using X-ray absorption spectra measured in total electron yield (TEY) mode.

To relate the imaging data to the spectroscopy measurements, the peak intensities at the $L_2$ edge obtained from both datasets are compared by defining the peak ratio
\begin{equation}
\gamma = \frac{I_{\mathrm{spectroscopy}}}{I_{\mathrm{image}}}.
\label{eq:si_eq_4}
\end{equation}
Using this ratio, the absorption maximum expected for the transmission imaging experiment is estimated as
\begin{equation}
A_{\mathrm{max}}^{\mathrm{image}} = \frac{A_{\mathrm{max}}^{\mathrm{spectroscopy}}}{\gamma}.
\label{eq:si_eq_5}
\end{equation}
The resulting $A_{\mathrm{max}}^{\mathrm{image}}$ is then used in Eq.~(\ref{eq:si_eq_3}) to quantitatively evaluate the XMCD obtained from the transmission experiment. The XMCD signal obtained by this procedure reaches approximately $1.5\%$ at the $L_2$ edge.

\section{Supplementary Simulations}

\begin{figure}[h]
\begin{center}
\includegraphics[width=0.8\columnwidth]{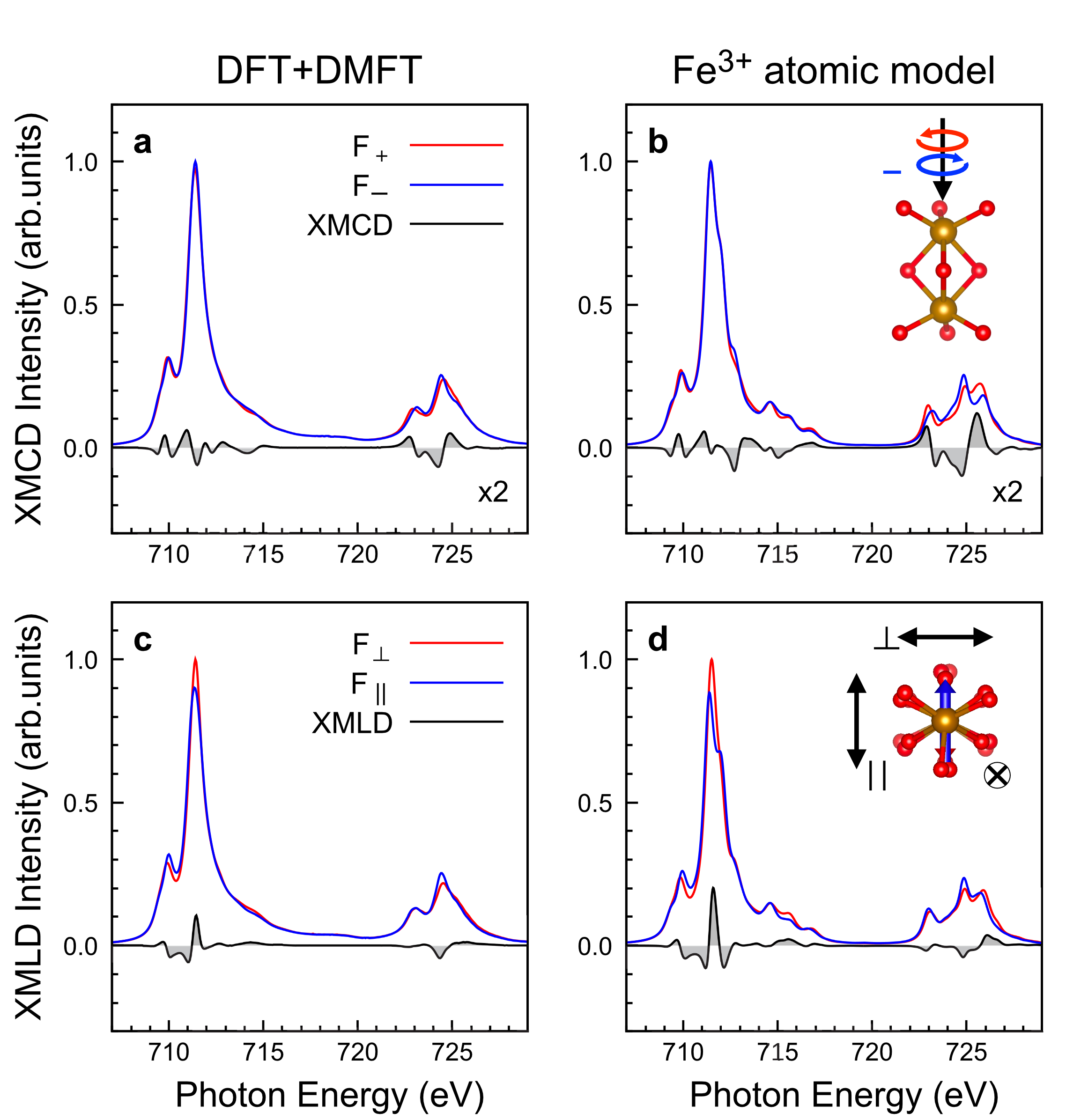}
\end{center}
\caption{Fe $L_{2,3}$-edge XMCD (top) and XMLD (bottom) spectra of $\alpha$-Fe$_2$O$_3$ calculated using the DFT+DMFT method (left) and an Fe$^{3+}$ atomic model (right). The X-ray geometries used in the calculations are illustrated in the insets of panels \textbf{b} and \textbf{d}.}
\label{si_fig_5}
\end{figure}

\begin{figure}[t]
\begin{center}
\includegraphics[width=0.5\columnwidth]{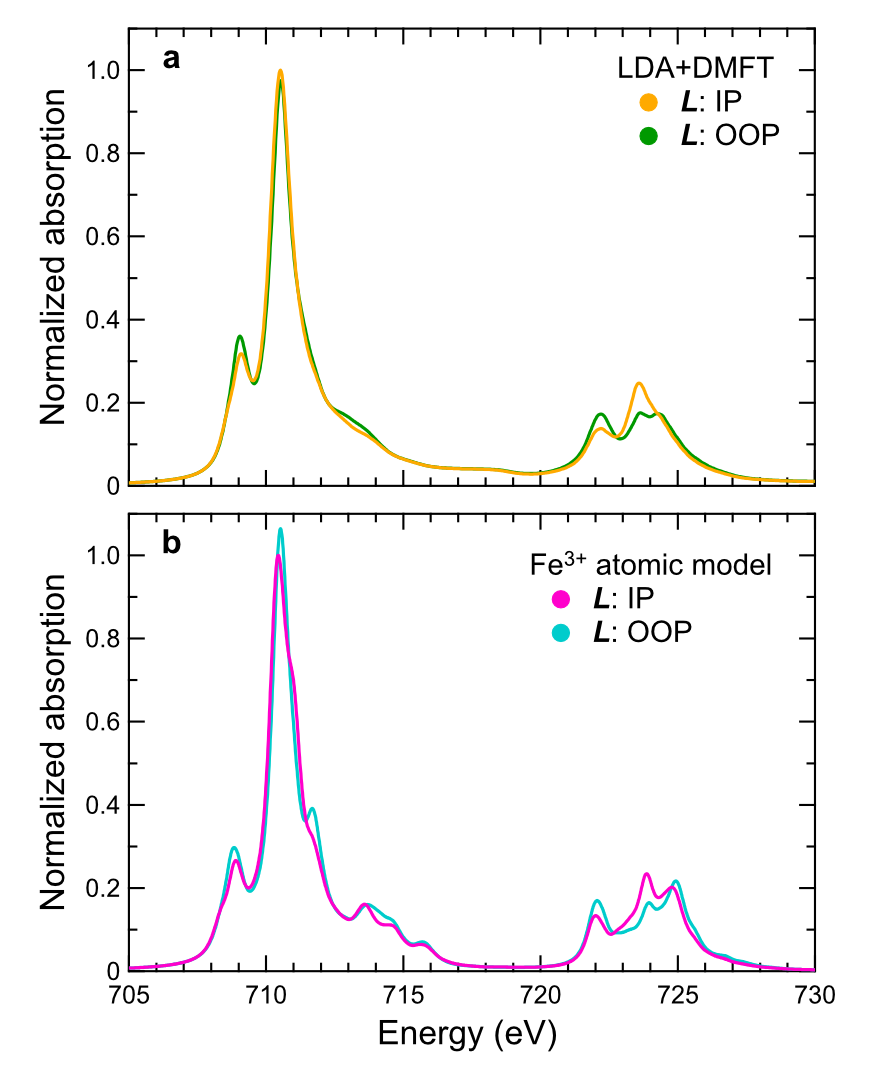}
\end{center}
\caption{Dependence of the XAS spectra on the $\mathbf{L}$ orientation (IP: in-plane, OOP: out-of-plane), calculated by (a) the DFT+DMFT method (same as Fig.~3\textbf{c} in the main text) and \textbf{b} the Fe$^{3+}$ atomic model.}
\label{si_fig_6}
\end{figure}

\begin{figure}[h]
\begin{center}
\includegraphics[width=0.8\columnwidth]{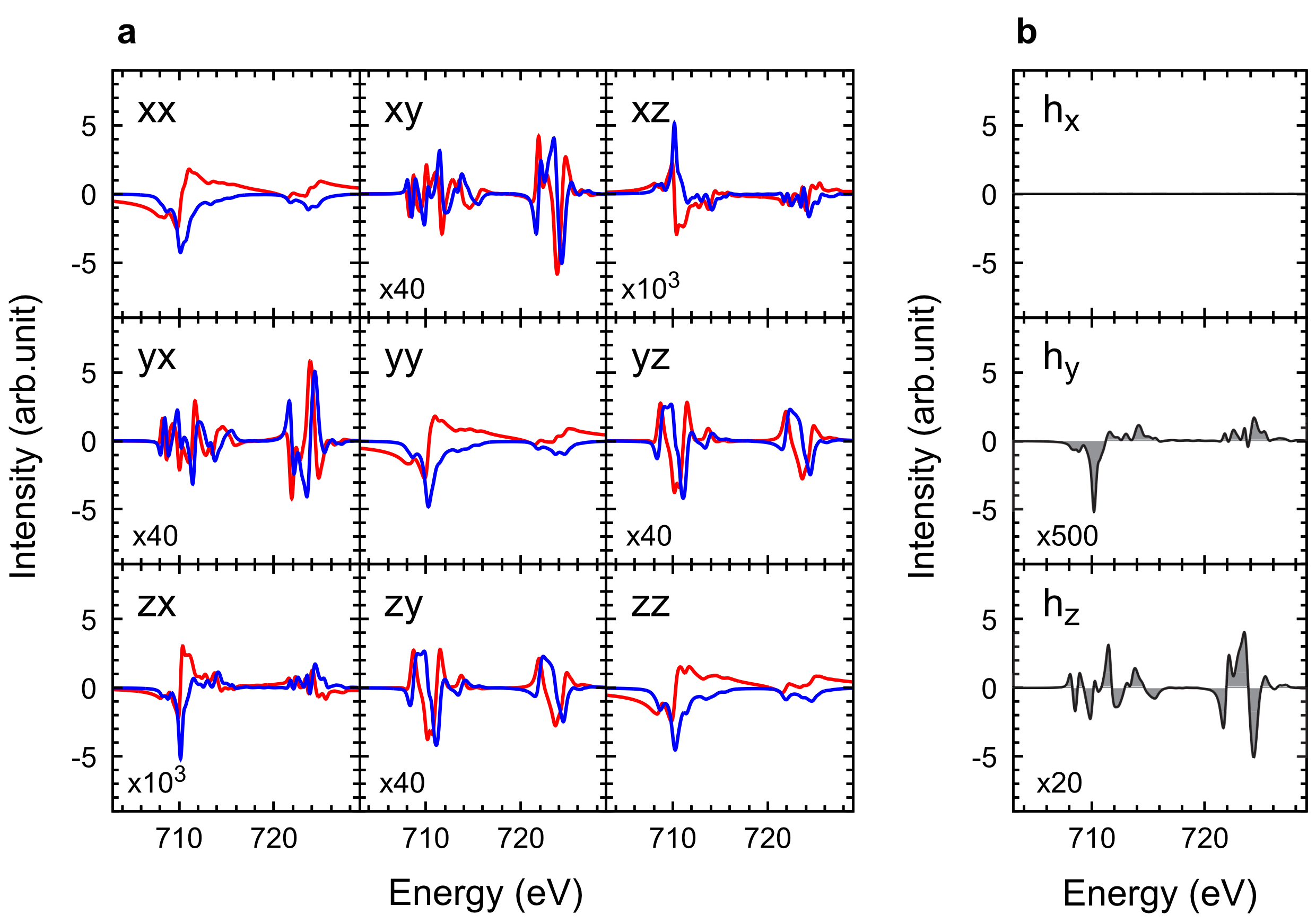}
\end{center}
\caption{\textbf{a} Real (red) and imaginary (blue) parts of the optical conductivity tensor at the Fe $L$ edge. \textbf{b} Imaginary part of the Hall vector $\mathbf{h}(\omega)$. The spectra are calculated using the Fe$^{3+}$ atomic model.}
\label{si_fig_7}
\end{figure}

\begin{figure}[t]
\begin{center}
\includegraphics[width=0.5\columnwidth]{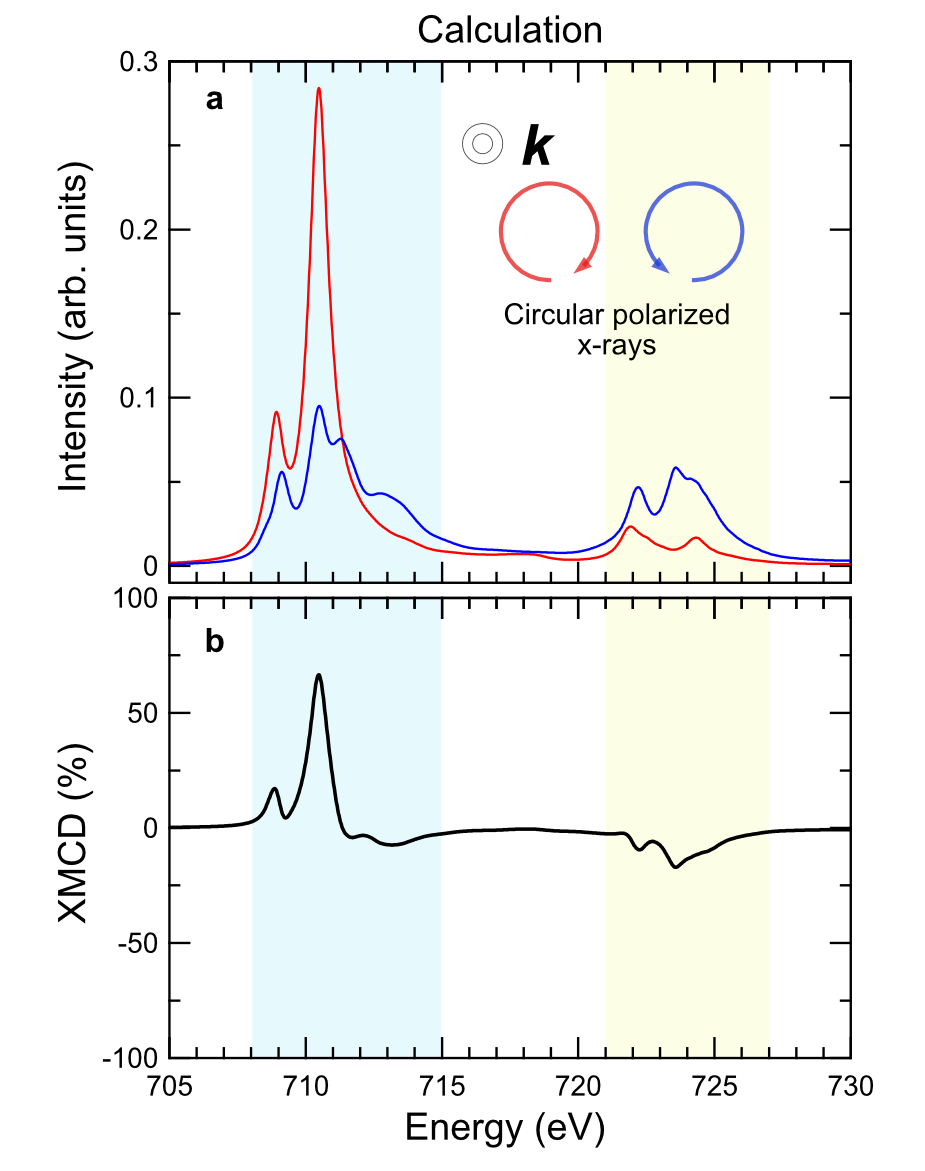}
\end{center}
\caption{\textbf{a} Fe $L$-edge XAS spectra calculated for circularly polarized light and \textbf{b} the corresponding XMCD spectra for a ferromagnetic configuration along the $c$ axis. Here the DFT+DMFT method is employed.}
\label{fig_theo_fm}
\end{figure}

Supplementary Figures~\ref{si_fig_5} and \ref{si_fig_6} compare the simulated Fe $L$-edge XMCD, XMLD and XAS spectra in the high-temperature altermagnetic phase calculated using the DFT+DMFT method with those obtained from an Fe$^{3+}$ atomic model. The crystal-field parameters in the atomic model are taken from a DFT calculation by constructing an effective Fe $d$-space model in which the O 2$p$ states are projected out. 
Since $\alpha$-Fe$_2$O$_3$\ is a good Mott insulator with localized Fe 3$d$ electrons, it is not surprising that the DFT+DMFT results and the atomic model yield similar absorption spectra. The broader high-energy feature around 715~eV in the DFT+DMFT spectrum, which shows improved agreement with the experimental data, is provided by the explicit inclusion of hybridization with the O 2$p$ bands. 
In the atomic model, a small molecular field is introduced to simulate the N\'eel state of this phase. 
A small in-plane component is included to account for the canting of the magnetic moment, whose amplitude is adjusted to reproduce the experimental magnetization of $8.0\times10^{-3}\,\mu_{\rm B}/{\rm Fe}$.
Supplementary Figure~\ref{si_fig_7}(a) shows the optical conductivity tensor at the Fe $L$ edge calculated using the atomic model, and the corresponding Hall vector $\mathbf{h}(\omega)$ is shown in Supplementary Fig.~\ref{si_fig_7}(b). The XMCD is given by $2\,\mathrm{Im}\,[\mathbf{h}(\omega)\!\cdot\!\hat{\mathbf{k}}]$, where $\mathbf{k}$ is the X-ray wave vector~\cite{Hariki24}. In the adopted coordinate system, the $x$, $y$, and $z$ axes correspond to the $a$, $m$, and $c$ axes shown in Fig.~1 of the main text, respectively. The in-plane components of the Hall vector are symmetry-allowed but negligibly small compared with the out-of-plane component $h_z(\omega)$. We have also checked that the presence or absence of the in-plane canted moment does not alter $h_z(\omega)$, confirming that it represents the altermagnetic contribution. Finally, Supplementary Fig.~\ref{fig_theo_fm} shows the XAS and XMCD spectra calculated using the DFT+DMFT method for a ferromagnetic configuration with magnetic moments aligned parallel to the $c$ axis. A large XMCD signal appears, as expected, but its frequency dependence is significantly different from that of the altermagnetic case. This is another demonstration of the fact that the observed XMCD spectrum is not due to any net magnetic moment
parallel to the $c$ axis, which is not observed but is symmetry allowed.


\clearpage

\bibliography{fe2o3}